%% file: main.tex
\documentclass[acmsmall]{acmart}

\usepackage{graphicx} 
\usepackage{cleveref}
\usepackage{subcaption}
\usepackage{listings}
\usepackage{wrapfig}
\usepackage{amsfonts}
\usepackage{amsmath}
\usepackage{tikz}
\usetikzlibrary[arrows.meta,bending]
\usetikzlibrary[positioning]

\usepackage{enumitem}

\usepackage[nofillcomment,linesnumbered,ruled,vlined,noend]{algorithm2e} 

\SetCommentSty{mycommfont} 

\usetikzlibrary{positioning,decorations.pathreplacing,fit}
\newcommand{\tikzmark}[2][]{%
  \tikz[remember picture,overlay,baseline=-.5ex] \node[#1] (#2) {};%
}
\newcommand*{\drawBrace}[4][0pt]{%
    \node[draw=none, fit={(#2) (#3)}, inner sep=0pt] (rectg) {};%
    \draw [decoration={brace,amplitude=0.3em},decorate,thick,mygray]%
      ([xshift=#1]rectg.north east) --%
      coordinate[right=1em, midway] (#2#3)
      ([xshift=#1]rectg.south east);%
    \node[right=-0.75em of #2#3] (#2#3-comment) {\textcolor{mygray}{#4}};
}%



\newcommand{\bnfdef}{\mathrel{::=}}
\newcommand{\bnfalt}{\mathrel{\mid}}

\title{Task-Based Tensor Computations on Modern GPUs}

\definecolor{keywordcolor}{rgb}{0.5,0,0.5}
\definecolor{textgray}{gray}{0.4}
\definecolor{mygray}{rgb}{0.5,0.5,0.5}
\author{Rohan Yadav}
\affiliation{%
    \institution{Stanford University}
    \city{Stanford}
    \state{California}
    \country{USA}
}
\authornote{This work was done while Rohan Yadav and Alex Aiken were at NVIDIA.}
\email{rohany@cs.stanford.edu}
\author{Michael Garland}
\affiliation{%
    \institution{NVIDIA}
    \city{Santa Clara}
    \state{California}
    \country{USA}
}
\email{mgarland@nvidia.com}
\author{Alex Aiken}
\affiliation{%
    \institution{Stanford University}
    \city{Stanford}
    \state{California}
    \country{USA}
}
\authornotemark[1]
\email{aiken@cs.stanford.edu}
\author{Michael Bauer}
\affiliation{%
    \institution{NVIDIA}
    \city{Santa Clara}
    \state{California}
    \country{USA}
}
\email{mbauer@nvidia.com}

\definecolor{todocolor}{rgb}{0.8,0,0}
\definecolor{editcolor}{rgb}{0,0,0.8}
\newcommand{\TODO}[1]{{\color{todocolor}#1}}
\newcommand{\IGNORE}[1]{}

\newcommand{\name}{Cypress}
\newcommand{\cublas}{cuBLAS}
\newcommand{\cudnn}{cuDNN}

\begin{abstract}

Domain-specific, fixed-function units are becoming increasingly
common in modern processors.
As the computational demands of applications evolve, the capabilities
and programming interfaces of these fixed-function units
continue to change.
NVIDIA's Hopper GPU architecture contains multiple fixed-function units per compute unit, including an asynchronous data movement 
unit (TMA) and an asynchronous matrix multiplication unit (Tensor Core).
Efficiently utilizing these units requires a fundamentally
different programming style than previous architectures; 
programmers must now develop warp-specialized kernels that orchestrate
producer-consumer pipelines between the asynchronous units.
To manage the complexity of programming these new architectures,
we introduce \name{}, a task-based programming model with sequential semantics.
\name{} programs are a set of designated functions called \emph{tasks}
that operate on \emph{tensors} and are free of communication
and synchronization.
\name{} programs are bound to the target machine through a
\emph{mapping} specification that describes where tasks should run and
in which memories tensors should be materialized.
We present a compiler architecture that lowers \name{} programs into
CUDA programs that perform competitively with expert-written codes.
\name{} achieves 0.88x-1.06x the performance of \cublas{} on
GEMM, and between 0.80x-0.98x the performance of the currently best-known
Flash Attention implementation while eliminating all aspects of explicit data movement and asynchronous computation from application code.

\end{abstract}

\setcopyright{cc}
\setcctype{by}
\acmDOI{10.1145/3729262}
\acmYear{2025}
\acmJournal{PACMPL}
\acmVolume{9}
\acmNumber{PLDI}
\acmArticle{163}
\acmMonth{6}
\received{2024-11-12}
\received[accepted]{2025-03-06}

\begin{document}

\maketitle

\section{Introduction}


To continue delivering performance and efficiency improvements for important applications, modern computing architectures are becoming increasingly heterogeneous.
Heterogeneity is now common even \emph{within} a processor like a GPU.
A modern GPU contains several domain-specific fixed-function units for accelerating data movement~\cite{hopper-arch},
matrix multiplication~\cite{volta-arch}, 
texture interpolation~\cite{texture-caches}, and ray tracing~\cite{turing-arch}.
As the demands of applications evolve, the characteristics of these fixed function units, their capabilities, and programming interfaces can rapidly change. 

An extreme example of the rate of architectural change
and its impact on programming interfaces can be seen in the explosive growth of 
NVIDIA's Tensor Cores~\cite{volta-arch}, which perform 
matrix-multiply-accumulate (MMA) operations.
Tensor Cores have grown dramatically in six years
, from computing 8x8x4 multiplications
in the Volta~\cite{volta-arch} architecture to 64x256x16 multiplications in
the latest Hopper architecture~\cite{hopper-arch}.
As the capabilities of Tensor Cores have increased, their programming model underwent significant revisions.
On the Volta and Ampere~\cite{ampere-arch} architectures, groups of 8 and 32 threads
dispatch matrix multiplications onto the Tensor Cores, which interact with the register file.
In contrast, Hopper's single Tensor Core 
per streaming multiprocessor (SM) 
requires groups of 128 threads to collectively issue a matrix multiplication on tiles of data split across
both the register file and shared memory.
%
To keep Hopper's Tensor Cores fed with data, an asynchronous data movement unit called 
the Tensor Memory Accelerator~\cite{hopper-arch} (TMA) must be
used.

Achieving peak performance using this pair of functional units requires writing 
complex pipelines that 
manage synchronization between asynchronous copies and computation~\cite{flash-attention-3, fa2-hopper, cutlass-gemm}, a fundamentally different 
programming style than in prior architectures (\Cref{sec:background}).
Managing the asynchrony required to efficiently utilize these
latest functional units adds a new level of difficulty to developing 
high performance GPU code.
While kernel libraries can be provided by vendors, the development
of new kernels and the development of
the kernel libraries themselves directly face the challenges of programming
with asynchronous fixed-function units.
%
State-of-the-art programming systems that target Hopper either push the brunt of managing communication
and synchronization to the programmer (such as kernels written using extensions of CUTLASS~\cite{cutlass} or ThunderKittens~\cite{thunderkittens}) or
exclude these concepts from the programming model and rely on the compiler to infer all such details (such as Triton~\cite{triton}).
Manual management of data movement and synchronization is error-prone even for expert
programmers, leading to data races or sub-optimal performance when opportunities for overlap
are not exploited.
More automatic systems ease some of these burdens for the programmer, but require the compiler to make
heuristic decisions about how to map the computation and data onto the physical machine.
As we show in \Cref{sec:evaluation}, automated systems like Triton can produce sub-optimal performance when
compared to expert-written code.
We believe that programming models for these emerging accelerators must evolve
to meet dichotomous goals: making it easy to write correct code in the face of increasingly complex architectures while simultaneously delivering low-level control over crucial decisions for achieving peak performance.

To this end we present Cypress, a task-based programming model
to manage the asynchrony and heterogeneity in emerging GPU architectures.
%
The Cypress programming model aims to find a middle ground between prior approaches
by automating difficult aspects like explicit data movement and synchronization,
but gives the user control over performance-critical decisions like how
a computation is decomposed and mapped onto the target machine.
\name{} targets tensor algebraic computations that leverage
asynchronous fixed-function units like the Tensor Core and TMA.
%
%
A Cypress program is organized into two components: a \emph{logical description} of the computation
and a \emph{mapping specification} that binds the computation to the physical machine.
The logical description of a Cypress program 
describes the computation sequentially in terms of designated functions called \emph{tasks} that operate on first-class \emph{tensors}.
%
%
%
%
The mapping specification of a Cypress program enables users to dictate 
on which groups of threads tasks should execute (e.g., thread blocks
or warps), and in which memories tensors should be placed  (e.g., global memory or shared memory).
The Cypress compiler combines the logical computation with the mapping to infer where implicit parallelism can be exploited and where data movement must be inserted for correctness to produce
an aggressive static schedule that fully leverages available functional units.  
%
We envision that a system like \name{} could be used in the development of new kernels that are not
supported by existing vendor libraries, or used as a tool to develop kernels within the
vendor libraries themselves.


We implement a prototype of the \name{} compiler as a source-to-source translator that combines the logical computation and
mapping specification to generate CUDA C++ for the data movement and matrix multiplication accelerators
available on NVIDIA Hopper GPUs.
%
%
Our prototype compiler is implemented in MLIR~\cite{mlir} and 
lowers away task-based abstractions into {\em warp-specialized} code \cite{cuda-dma} structured similar to CUTLASS main loops.
The specific contributions of this work are:
\begin{itemize}
    \item A programming model with sequential semantics for targeting GPUs with
    asynchronous fixed-function units through task-based abstractions.
    \item A compiler architecture that lowers task-based programs to GPU code that achieves near peak performance. 
\end{itemize}

To evaluate \name{},
we implement a variety of linear algebra kernels
such as GEMM and Fused Multi-Headed Attention.
\name{} performs competitively with hand-tuned implementations in 
\cublas{}, \cudnn{}, CUTLASS and ThunderKittens,
achieving between 0.80x-1.09x of their performance.
We also show that first-class asynchrony and control over mapping enables kernels developed in \name{}
to outperform Triton by 0.99x-2.18x, as Triton
often struggles to effectively utilize the TMA
and its heuristics routinely place tensors in suboptimal locations in memory. 

%

\section{Background}\label{sec:background}

Modern GPUs are programmmed using a SIMT (Single Instruction, Multiple Threads) programming model: every thread executes the same sequence of instructions.  Kernels are usually written in a bulk-synchronous style, where all threads in a block periodically synchronize with each other at a barrier. Linear algebra kernels, specifically, are often statically parameterized with tile sizes and software pipelining depths~\cite{blis,atlas}. 
%
%
%
%
%
With the advent of deep learning, GPU architects introduced fixed-function matrix multiplication units to continue increasing GEMM performance.
%
These units have grown to perform asynchronous
bulk computations over large tiles of data,
and their evolution has dramatically
affected the structure of efficient GEMM programs.
An inflection point occurred between the NVIDIA Ampere~\cite{ampere-arch}
and Hopper~\cite{hopper-arch} architectures, where an optimized Hopper GEMM
kernel requires a substantially different structure than an optimized Ampere GEMM kernel.
To motivate the need for systems like Cypress, we present psuedocode
for efficient GEMM kernels on Ampere and Hopper in \Cref{fig:ampere-hopper-gemm},
adapted from CUTLASS~\cite{cutlass} implementations.
%
Each example
%
shows the computation that executes on an
individual streaming multi-processor (SM) of the target GPU, which is responsible
for computing a tile of the output matrix $C = A \times B$ of size \texttt{T\_M}
by \texttt{T\_N}.

\subsection{Programming Ampere}

\begin{figure}
\begin{subfigure}[h]{0.5\linewidth}
\begin{lstlisting}
def mma_device(gC, gA, gB):
  # Shared memory tiles of A and B.
  _shared_ sA[T_M,T_K], sA_next[T_M,T_K]
  _shared_ sB[T_K,T_N], sB_next[T_K,T_N]
  for k in range(0, gA.shape[1] / T_K):(*\label{fig:ampere-kloop}*)
    # All threads prefetch tile of global memory.
    copy(tile(gA, (blk_x(), k+1)), sA_next)(*\label{fig:ampere-smem-prefetch}*)
    copy(tile(gB, (k+1, blk_y())), sB_next)
    # Allocate register fragments for 
    # thread-local pieces of A and B.
    rA, rA_next = allocFragA(), allocFragA()
    rB, rB_next = allocFragB(), allocFragB()
    copy(tile(sA, 0), rA)
    copy(tile(sB, 0), rB)
    # Block the shared memory tiles into 
    # register-sized chunks.
    for kk in range(0, T_K / BLK_K):
      # Prefetch shared memory into registers.
      copy(tile(sA, kk+1), rA_next)(*\label{fig:ampere-rmem-prefetch}*)
      copy(tile(sB, kk+1), rB_next)
      # Fully unrolled thread-level gemm.
      gemm(accum, rA, rB)
      # Swap rA, rB with rA_next, rB_next
      swap(rA, rA_next)
      swap(rB, rB_next)
    # Swap sA, sB with sA_next, sB_next
    swap(sA, sA_next)
    swap(sB, sB_next)
  # Flush the accumulator to global storage. 
  copy(accum, tile(gC, blk_x(), blk_y()))
\end{lstlisting}
\caption{Ampere GEMM main loop~\cite{cutlass-ampere-gemm}.}
\label{fig:ampere-gemm}
\end{subfigure}\hfill%
\begin{subfigure}[h]{0.5\linewidth}
\begin{lstlisting}
def mma_device(gC, gA, gB):
  # Shared memory tiles of A and B.
  _shared_ sA[T_M,T_K,PIPE], sB[T_K,T_N,PIPE]
  _shared_ sC[T_M,T_N] # Staging buffer for C.
  barrier prod[PIPE], cons[PIPE], copyout
  if tid() >= 128:
    # One warp (32 threads) performs copies.
    for k in range(0, gA.shape[1] / T_K):(*\label{fig:dmawarp-start}*)
      if k >= PIPE:
        wait(cons[k%PIPE]) # Wait for consumer.
      if tid() == 128:
        # Only 1 thread should invoke the TMA.
        TMA_load((*\label{fig:tmaload}*)
          prod[k % PIPE], # Completion barrier.
          tile(gA, (blk_x(), k)) -> sA[_,_,k%PIPE],
          tile(gB, (k, blk_y())) -> sB[_,_,k%PIPE])
    wait(copyout)
    if tid() == 128:
      TMA_store(sC -> tile(gC, blk_x(), blk_y()))(*\label{fig:dmawarp-end}\label{fig:tmastore}*)
  else:
    # Four warps (128 threads) perform GEMMs.
    for k in range(0, gA.shape[1] / T_K):(*\label{fig:computewarp-start}*)
      wait(prod[k%PIPE]) # Wait for TMA.(*\label{fig:waittma}*)
      # All 128 threads concurrently call wgmma.
      warpgroup_sync()
      # Invoke Tensor Core and wait for result.
      wgmma(accum, sA[_,_,k], sB[_,_,k])(*\label{fig:wgmma-call}*)
      warpgroup_wait()(*\label{fig:waitwgmma}*)
      arrive(cons[k%PIPE]) # Notify DMA warp.(*\label{fig:notify-dma-warp}*)
    # Copy from registers to the staging buffer.
    copy(accum, sC)
    syncthreads()
    arrive(copyout)(*\label{fig:computewarp-end}*)
\end{lstlisting}
\caption{Hopper GEMM main loop~\cite{cutlass-hopper-gemm}.}
\label{fig:hopper-gemm}
\end{subfigure}
\caption{High-level GEMM computation structure on Ampere and Hopper GPUs.}
\label{fig:ampere-hopper-gemm}

\end{figure}

\Cref{fig:ampere-gemm} presents psuedocode for a high performance GEMM implementation targeting the Ampere architecture.
At a high level, the implementation allocates shared memory buffers for tiles of the $A$ and $B$
matrices, and iterates over tiles in the $K$-reduction dimension of the computation (Line~\ref{fig:ampere-kloop}).
The implementation executes a multi-stage software pipeline, prefetching data from the global
high-bandwidth memory into the SM's local shared memory (Line~\ref{fig:ampere-smem-prefetch}), and then gathering data from
the local shared memory into each thread's register file (Line~\ref{fig:ampere-rmem-prefetch}).
A sophisticated implementation in CUTLASS~\cite{cutlass} or cuBLAS would use specialized data movement
instructions~\cite{graphene} that load matrices from shared memory into the registers, 
invoke the tensor core at the warp level, and ensure that instructions are interleaved 
in a manner that keeps all floating point and load/store units busy at the same time.
%

\subsection{Programming Hopper}

While the Ampere GEMM implementation is a fairly traditional bulk-synchronous CUDA program, the
Hopper implementation in \Cref{fig:hopper-gemm} has pervasive asynchrony, requiring a fundamentally
different approach.
%
%
Although each SM still iterates through tiles in the K-reduction dimension, for Hopper, almost all data movement is offloaded asynchronously to the TMA (Lines~\ref{fig:tmaload} and~\ref{fig:tmastore}), and the bulk of the GEMM computation is offloaded asynchronously to the Tensor Core (Line~\ref{fig:wgmma-call}).
The high throughput of both the TMA and the Tensor Cores mandates the use of deep
pipelines (controlled by the \texttt{PIPE} variable) to avoid exposing latency.
%
%
Meanwhile, the threads of the SM must manage the dependencies between both the fixed function units by periodically synchronizing with them while simultaneously performing their own computations (Lines~\ref{fig:waittma} and~\ref{fig:waitwgmma}).
It is challenging to maintain a schedule that manages the
competing concerns of ensuring functional correctness and avoiding
unnecessary synchronization that hinders performance.
%

To further complicate matters, achieving peak performance mandates
using {\em warp specialization}~\cite{cuda-dma} to decouple 
data loading from computation. 
%
%
%
With warp specialization for Tensor Cores, \emph{warps} (groups of 32 hardware threads) are specialized so that there is a 
data-movement warp (DMA) that exclusively issues work to the TMA (Lines~\ref{fig:dmawarp-start}--\ref{fig:dmawarp-end}) and a group
of four warps (henceforth referred to as a \emph{warpgroup}) that exclusively issue work to the Tensor Core (Lines~\ref{fig:computewarp-start}--\ref{fig:computewarp-end}).  The choice of four warps in a warpgroup is mandated by the machine--every Tensor Core operation must be cooperatively launched by exactly four warps (128 threads).  
%
The DMA warp invokes the TMA (from a single thread) to load tiles into shared memory, and then notifies the compute warpgroup via the \texttt{prod} barrier when tiles are available.
When the \texttt{prod} barrier is triggered, the
compute warpgroup collectively issues the matrix multiplication
to the Tensor Core (Line~\ref{fig:wgmma-call}).
The operands matrices of the matrix multiplication are partitioned
across each of the 128 threads in the warpgroup, split across
the registers and the shared memory.
Then, all of the 128 threads execute the \texttt{wgmma}
instruction at the same time to invoke the Tensor Core.
After explicitly waiting for the Tensor Core to
complete, the compute warpgroup notifies the
DMA warp via the \texttt{cons} barrier so that
shared memory buffers can be reused (Line~\ref{fig:notify-dma-warp}).
%
Warp specialization also improves efficiency by enabling
the registers of the data movement warp to be used by
the compute warps to store larger accumulators in the register file.
%
%
Taken in aggregate, these differences signal that for Hopper and later GPUs, kernels can no longer be structured in the traditional bulk-synchronous style, and instead must be organized differently to cope with the significant challenges posed by new and changing functional units.

\section{Programming Model}\label{sec:programming-model}

If these trends become common, asynchronous fixed-function units will become increasingly
powerful, and programming models will need to adapt to aid programmers in managing this level of pervasive asynchrony.
%
The \name{} programming model aims to provide programmers with
tools to grapple with these growing complexities while still enabling peak performance.
Similar to other task-based programming models~\cite{legion, sequoia, parsec, starpu, ray}, Cypress programs have sequential semantics, which simplifies reasoning about correctness and ensures that common bugs involving races and data coherence cannot manifest.
However, unlike most task-based models (except Sequoia~\cite{sequoia}, see Section~\ref{sec:related-work}), the Cypress model is designed to be amenable to a fully static analysis so programs can be statically parallelized, scheduled, and optimized without incurring runtime overheads that would inhibit peak performance.
A Cypress program consists of a logical description of the computation in terms of \emph{tasks} that operate on
collections of data called \emph{tensors}, where all tasks appear to execute in sequential program order (Section~\ref{sec:log-desc}).
Separately, the mapping specification of a Cypress program (Section~\ref{sec:mapping-spec}) binds the tasks and tensors onto a target machine.
\name{} is designed such that choice of mapping decisions can only affect performance, not correctness.
%
The Cypress compiler (\Cref{sec:compilation}) is responsible for inferring parallelism, inserting all data movement required to maintain coherence, and enforcing the 
necessary synchronization required to maintain the sequential semantics of the source program under the mapping description.
To motivate the need for many of the programming model's features we begin with a
discussion of the machine model.

\subsection{Machine Model}

\begin{wrapfigure}{r}{0.26\textwidth}
    \centering
    \includegraphics[width=0.23\textwidth]{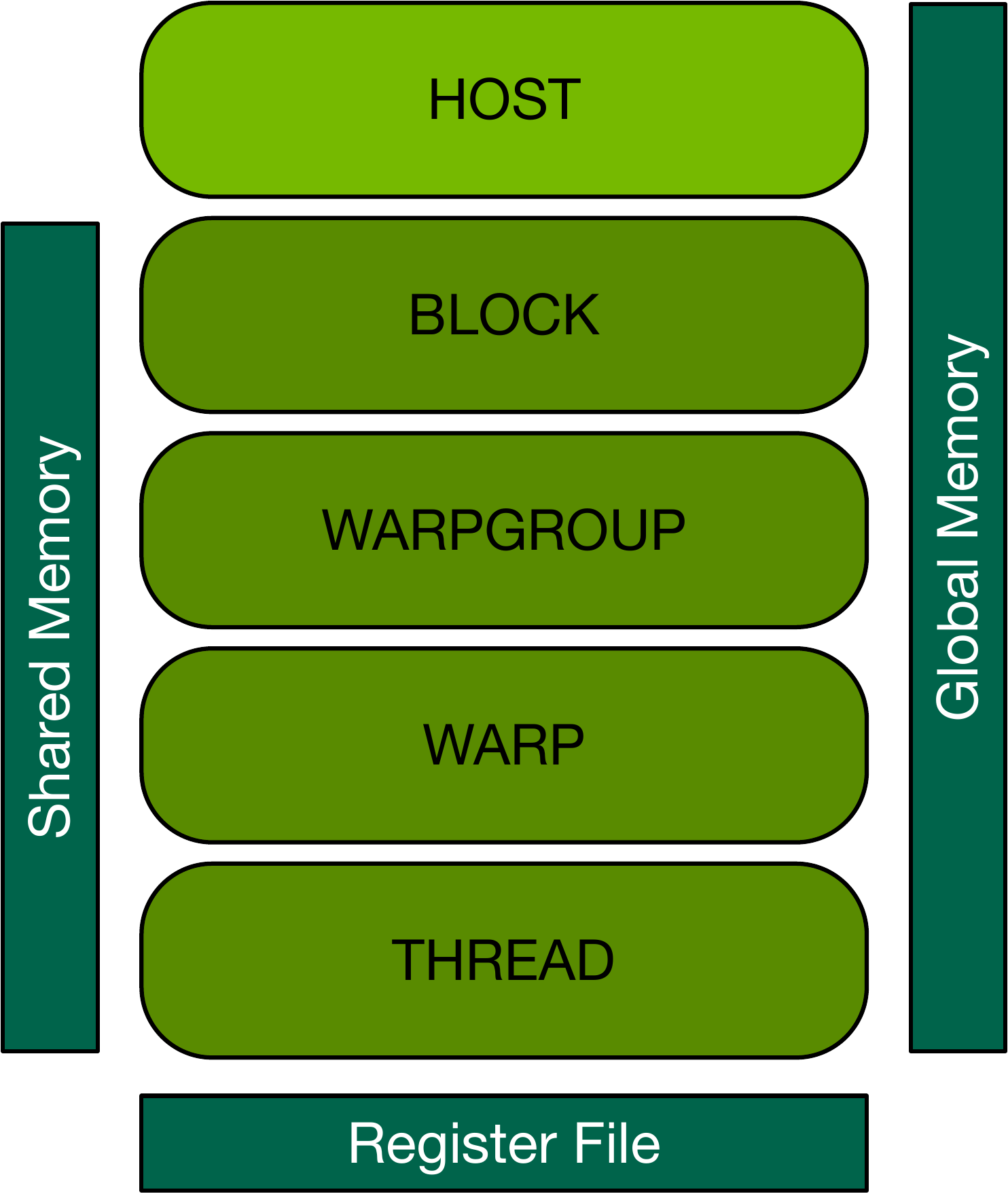}
    \caption{H100 Machine Model}
    \label{fig:h100-machine-model}
\end{wrapfigure}

\name{} describes machines to the programmer using a 
hierarchical model that
captures the locality in modern architectures.
A machine is described by specifying
one or more \emph{processor levels}, each of which corresponds to
a logical processor on the target machine, corresponding to
thread groupings that have semantic meaning in the target machine.
The machine also describes the available memories and which
processor level(s) can access each memory.
A machine description for the Hopper GPU is shown
in \Cref{fig:h100-machine-model}.
%
%
%
%
The Hopper machine description has multiple processor levels,
from the host that launches GPU kernels down to the
individual threads on the device.
Rather than representing Tensor Cores explicitly, 
we introduce the warpgroup level to describe a group of threads 
capable of initiating work on a Tensor Core.
The Hopper machine description has the usual 3 kinds of memories supported by CUDA, each visible from different processor levels.
The global memory is visible to all processors, while the shared memory is only
visible to an individual block and below, and the register file is thread local.
We expect this framework for describing machines to be robust at capturing salient architectural details in the face of continued GPU evolution by adding new memories and processor levels.
For example, the newly released NVIDIA Blackwell architecture features a larger tensor core shared by pairs of SMs and a new
memory kind to store tensor core output into.
Such features could be captured by additional hierarchy,
new processor and new memory kinds in the \name{} machine model.
A version of \name{} that targets a specific GPU architecture would
provide the programmer a machine model for that GPU.
The hierarchical nature of the processors in the machine model also provides a conceptual target for programmers to consider when structuring their programs and expressing mapping specifications.

\subsection{Logical Description}\label{sec:log-desc}

\begin{figure}
\begin{minipage}{.65\textwidth}
\input{fe-syntax}
\captionof{figure}{Abstract syntax for \name{} programs.}
\label{fig:frontend-grammar}
\end{minipage}\hfill%
\begin{minipage}{.30\textwidth}
\centering
\includegraphics[width=0.95\textwidth]{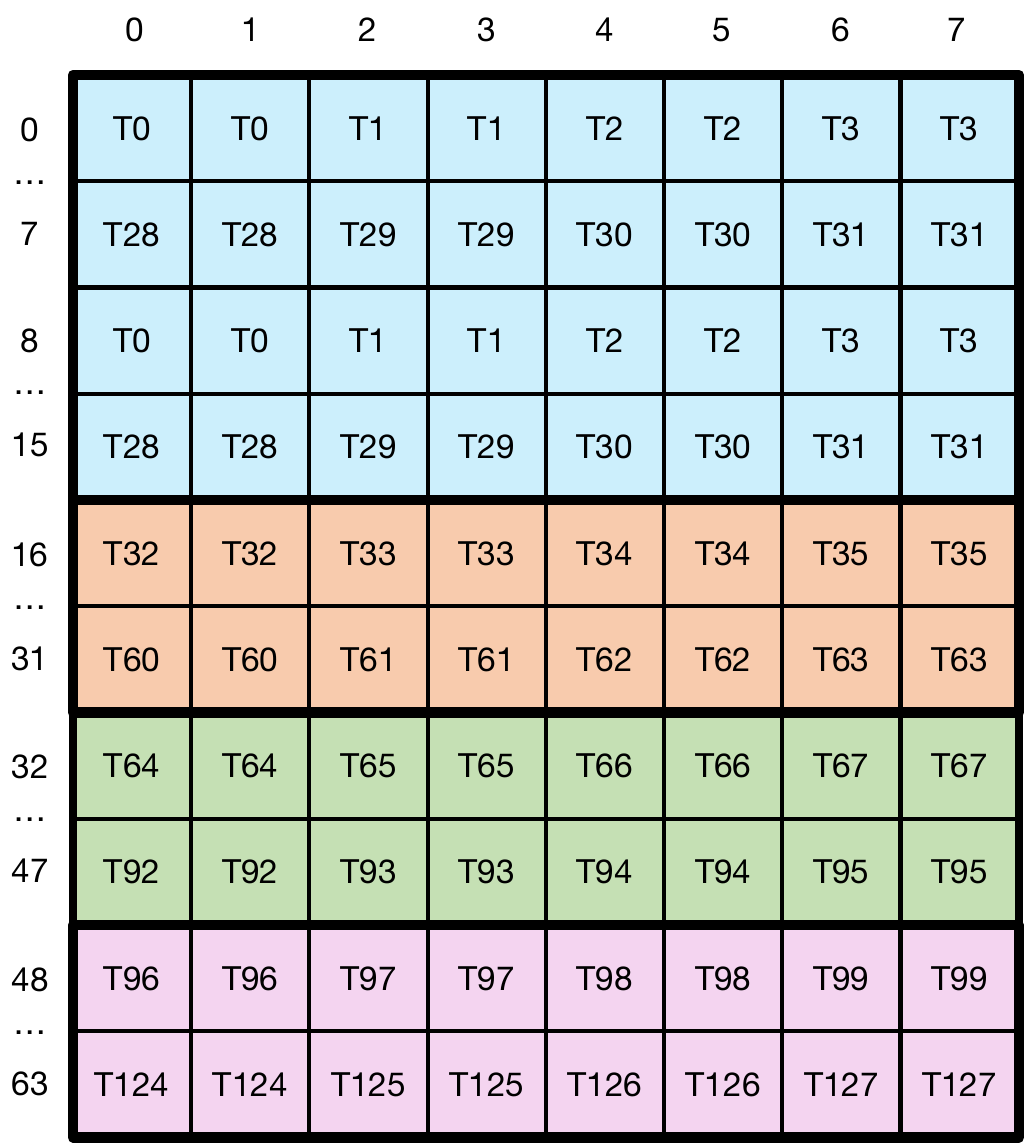}
\captionof{figure}{Output matrix layout in registers for M=64,N=n*8 warpgroup matrix multiplication instruction (adapted from NVIDIA PTX documentation).}
\label{fig:wgmma-acc-reg-layout}
\end{minipage}

\end{figure}

The first component of a Cypress program is the \emph{logical description},
which expresses the computation in terms of tasks and tensors.
Tasks and tensors are described without referencing
physical memory addresses or data movement mechanisms, so the same task implementations can be
potentially reused both at multiple levels within a target machine and across machines.
%
The logical description of \name{} programs is defined in the abstract
syntax in \Cref{fig:frontend-grammar}.
%

\paragraph{Tasks.}
Computations in \name{} are described as \emph{tasks}. 
A task is a designated function that may have multiple \emph{variants},
which are different implementations that may target
different processors or employ different algorithms.
Each variant of a task must have the same function signature.
%
%
Tasks are parameterized using \emph{tunable} values, which are statically 
specialized to values provided in the mapping specification.
Different variants of a task may require different tunable values to be specified.
Task variants may either be \emph{inner} or \emph{leaf} variants.
%
%
Inner task variants may launch \emph{sub-tasks} to express hierarchical
decomposition of the computation.
%
Inner task variants are limited to performing only scalar computations, partitioning
tensors (discussed later in this section) and launching sub-tasks --- inner
variants may not access the data within tensors or invoke external functions.
Leaf task variants may not launch sub-tasks, but may directly access tensor data
and perform arbitrary computation, such as invoking arbitrary C++ functions (using 
the \textsf{call-external} construct).

Inner task variants may launch sub-tasks in multiple ways, depending on the desired
algorithmic choice.
First, inner task variants may launch a sequential group of tasks
over an iteration domain
using the \textsf{srange} operation, where each task executes in order.
Inner tasks may alternatively use the \textsf{prange} operation to
launch a group of parallel tasks.
As discussed shortly, the tasks launched with \textsf{prange} must not perform aliasing 
writes to the same (parts of) tensors.
While tasks launched by \textsf{prange} execute in parallel, the sequential semantics
ensure that the effects of execution are the same as if
\textsf{srange} was used.
%
%
%
Finally, single tasks may be launched inline.
%
These three constructs for launching subtasks enable programmers to decompose computations 
differently to match the capabilities of different hardware resources.
In all cases, the concrete variant to be dispatched at a task launch site is controlled by the mapping specification, allowing customization to a target machine.

\paragraph{Tensors.}
\name{} has a first-class representation of data collections as
\emph{tensors}, or multi-dimensional arrays.
Tasks operate on tensors and accept both tensors and scalars as arguments.
Every task declares the effects it will have on its argument tensors
by specifying the \emph{privilege} with which each tensor will be used.
%
Tasks must respect the declared privilege when accessing a tensor or launching
sub-tasks.
For example, a task that requests the \textsf{read} privilege on a tensor may not launch
a sub-task that requests \textsf{write} privileges on the same tensor.
%
%
The privileges that tasks take on their argument tensors impact the optimizations and 
transformations that the \name{} compiler can perform.


Tensors may be \emph{partitioned} into pieces to decompose the data with the decomposed
computation.
Partitions of tensors are created through a set of \emph{partitioning operators} 
(similar to Diffuse~\cite{legate-kernel-fusion}) that describe the mechanism of partitioning.
These partitioning operators create subsets of a tensor, each of which is also a tensor.
The sub-tensors created by partitioning operators need not contain contiguous sets of 
elements in the original tensor as each sub-tensor is given a new compacted, origin-based coordinate system.
Sub-tensors may then be extracted from partitions with the indexing operation 
and used as arguments to task launches, expressing the data required by 
each sub-computation.

There are currently two partitioning operators, both of which will be used for implementing
the high performance GEMM in \Cref{sec:full-cypress-gemm}.
The \textsf{blocks} partitioning operator performs a tiling-based
partition, partitioning a tensor into blocks of a given size.
The \textsf{mma} partitioning operator abstracts the tensor
partitioning required for a Tensor Core.
\Cref{fig:wgmma-acc-reg-layout} contains an example of the partitioning for the output $C$ matrix of a Hopper Tensor Core matrix multiply  (note the partition is logically non-contiguous).
%
%
The Hopper Tensor Core expects the rows of the output matrix
to be partitioned into groups of 16 across the four warps in the 
warpgroup (shown by the coloring of \Cref{fig:wgmma-acc-reg-layout}), and the
columns of the output to be swizzled across the threads in each warp.
The swizzling pattern is repeated across each group of 8 rows (shown
only for the first warp).
%
The matrix elements that thread $i$ in the warpgroup holds are denoted by the label $Ti$
in \Cref{fig:wgmma-acc-reg-layout}.
%
%
This partitioning pattern is replicated across the number of columns available in the
MMA instruction --- for example, the same pattern would be used for columns 8-15 and 16-23
in the 64x24x16 MMA instruction.
These partitioning strategies are mandated by the architecture to
avoid resource conflicts when accessing the shared memory or register file, and
will vary across accelerator instruction variants.

The partitioning strategies for a GEMM illustrate the necessity of
multiple variants of a task: the desired decomposition of both computation and data
can be different at each processor level to interact well with the hardware.
%
For example, the GEMM computation (\Cref{sec:full-cypress-gemm}) will use 
block-based partitioning strategies to decompose the data onto each SM and 
warpgroup, and then architecture-specific partitioning strategies to target
the Tensor Core.
Even within the Tensor Core partitioning, different strategies are used at the warp and thread levels,
affecting the partitioning strategies for tensors co-partitioned with tensors intended for consumption
by the Tensor Core.


\subsection{Mapping Specification}\label{sec:mapping-spec}

\input{gemm-example}

The second component of a \name{} program is a \emph{mapping specification} that dictates
how to bind a program to a particular machine description.
The mapping specification allows for control over performance sensitive decisions, and
the \name{} compiler ensures that mapping decisions do not affect correctness.
The mapping specification statically instantiates a tree of tasks by determining which task variants are used at each processor level and in which memories the tensors used by each task will be placed. 
A mapping specification is defined by a set of \textsf{task-mapping} objects,
defined in \Cref{fig:frontend-grammar}.
Each \textsf{task-mapping} object (called an \emph{instance}) corresponds to 
an instantiation of a specific task.
%
%
Each instance is identified with a name, which
may be referenced by other instances.
%
A instance declares several attributes: 1) the task variant the instance will use for execution, 2) the processor the task variant should execute on, 3) the memory where each tensor argument of the task should be placed (each tensor can be placed in a different memory), 4) the concrete values that should be bound to tunable values in the task variant, and 
5) the \textsf{task-mapping} instance to dispatch each launched child task to.
%
The mapping specification can also control processor-specific behavior, such as 
whether to warp-specialize or pipeline the execution of a particular task instance.
We provide examples of each of these additional controls in \Cref{sec:full-cypress-gemm}.
While not shown in \Cref{fig:frontend-grammar}, the mapping specification can
control other performance-sensitive attributes like the data layouts of
tensors.
Control over these data layouts can mitigate the impacts of hardware effects like bank conflicts in the shared memory.


Programmers often want to guarantee in a mapping specification that,
for either performance or capacity reasons, some tensors are never materialized in a particular memory but instead share a mapping with task instances further down the task tree.
For example, in a Hopper GEMM implementation, the output accumulator
matrix held by each thread block should never be materialized in its entirety.
Instead, the accumulator should only be materialized in a partitioned form among
the registers of each thread involved in the GEMM.
Such a constraint can be expressed using the \texttt{none} memory
option when mapping a tensor.
The Cypress compiler will report if any tensors mapped to the \texttt{none} memory were unable to satisfy this constraint during compilation, indicating that the user needs to change their partitioning or mapping strategy to achieve the desired outcome.


\subsection{Case Study: Hopper GEMM}\label{sec:full-cypress-gemm}

Having presented the logical description and mapping specification components
of the \name{} language, we now walk through a high performance Hopper GEMM
implementation developed in \name{}.
\Cref{fig:cypress-gemm} contains the GEMM implementation, where the previously
presented abstract syntax has been concretized in a Python embedded DSL.
The program describes the data and computation decompositions for a GEMM that
is hierarchically blocked for each processor level, where a different
task variant is used for each processor level.
The entrypoint of the computation is the \texttt{gemm\_host} root task instance which uses the task variant of the same name to execute on the \texttt{HOST} processor level.
\texttt{gemm\_host} breaks the output tensor $C$ into a set of tiles using the \texttt{U} and \texttt{V} tunables, and defines the panels of $A$ and $B$ needed to compute each output tile (\Cref{fig:cypress-gemm-tasks}, Line~\ref{fig:cypressgemm:hostpart}).
It then uses the \texttt{prange} construct to launch a group of parallel tasks, where each sub-task is dispatched to a \texttt{gemm\_block} instance and is responsible for computing a separate output tile of $C$.

The \texttt{gemm\_block} instances use a different task variant and therefore a different algorithm for GEMM than \texttt{gemm\_host}.
Inside a block (or SM), taking advantage of locality using the shared memory and 
register file is important for GEMM performance.
To do so, \texttt{gemm\_block} creates an accumulator tensor that will eventually be mapped
into the registers.
\texttt{gemm\_block} launches a task to zero-initialize the accumulator,
then uses \texttt{srange} to iterate over tiles of the $K$-reduction 
dimension and recursively perform GEMMs into the accumulator (\Cref{fig:cypress-gemm-tasks}, Line~\ref{fig:cypressgemm:srange}).
After processing all tiles, \texttt{gemm\_block} launches a task to copy
the accumulator to the output.
We elide the sub-trees for \texttt{clear} and \texttt{copy}, which
are invoked by \texttt{gemm\_block}.
The implementations and instantiations of these task variants share a similar structure to the remainder of the GEMM computation we will describe.
The mapping of \texttt{gemm\_block} also requests for warp specialization
of the task body, and for a 3-deep pipeline to execute the loop over tiles (\Cref{fig:cypress-gemm-mapping}, Lines~\ref{fig:cypressmapping:knob-ws} and~\ref{fig:cypressmapping:knob-pipeline}).

The GEMM is then decomposed onto the resources within an SM --- the warpgroups, warps and threads.
\texttt{gemm\_block} dispatches to the \texttt{gemm\_tile} variant (\Cref{fig:cypress-gemm-mapping}, Line~\ref{fig:cypressmapping:gemmtiledispatch}), which partitions the GEMM
row-wise between two warpgroups.
This is an optimization for large tile sizes that
splits the accumulator across multiple warpgroups to lower the per-thread register
usage, as each CUDA thread is limited to 255 registers.
The mapping creates two task instances for the warp-
and thread-level partitioning of operands for the Tensor Core (\Cref{fig:cypress-gemm-mapping}, Lines~\ref{fig:cypressmapping:inst1} and~\ref{fig:cypressmapping:inst2}).
Both task instances use the \texttt{gemm\_inner} variant, but
instantiated with different values for the tunable variables.
%
%
\Cref{fig:cypress-gemm-mapping} finishes instantiating the task tree with
the leaf task \texttt{gemm\_thread}, which executes on individual GPU threads.
%
Each GEMM invocation in Line~\ref{fig:cypressgemm:srange}
of \Cref{fig:cypress-gemm-tasks} is decomposed
into two groups of 128 leaf task invocations of
\texttt{gemm\_thread}.
Cypress does not perform analysis or optimization of the contents of 
leaf tasks, so users may invoke arbitrary CUDA C++ functions within
a leaf task's body.
%
\texttt{gemm\_thread} uses CuTe~\cite{cute} to dispatch to the desired
assembly instruction to invoke the Tensor Core.

\subsection{Discussion}

The separation of the logical description and the mapping specification enables the programmer to specify
performance-sensitive decisions without making invasive modifications to the application code.
One example of this flexibility is the control over data movement through mapping decisions.
The change in mapping from \texttt{GLOBAL} to \texttt{SHARED} between the 
\texttt{gemm\_block} and \texttt{gemm\_tile} instances (\Cref{fig:cypress-gemm-mapping}, Lines~\ref{fig:cypressmapping:globalmap} and~\ref{fig:cypressmapping:nonemap}) implies communication of 
tiles needed by \texttt{gemm\_tile} from global memory to shared memory, 
for which Cypress automatically uses the TMA (if available on the target GPU).
Another example is adding a level of hierarchy, which a user
might do when transitioning a program to use multiple warpgroups.
To perform this transition, a program that doesn't utilize the warpgroup level of the 
processor hierarchy would add additional task variants that target the warpgroup level,
and adjust the mapping specification to dispatch to these variants.
The existing task variants and mapping specifications would not need to be modified.

The separation provided by Cypress's programming model also allows for the algorithmic 
description of the problem to be separated from the implementation details required
to realize the algorithm on a physical device.
This aspect of the Cypress programming model places it in a middle ground between systems like CUTLASS that require users to manually manage low-level functional details and systems like Triton that hide most performance-sensitive mapping decisions from users.
Cypress offers the programmer low-level control over the algorithm and mapping decisions, but automates the implementation of the strategy for performance and guaranteed correctness.
As we show in \Cref{sec:evaluation}, this ability for the programmer to influence the compiler's decisions
can be important to achieving peak performance.
Finally, the Cypress programming model makes the partitions of data and their usages explicit
in the source program.
As code is modified or new functionality is introduced, it is the responsibility of the compiler
to continue to maintain an execution consistent with the sequential semantics of the program.
In contrast, program modification in a lower-level programming model can inadvertently result in bugs due to missing
synchronization or communication between different uses of partitioned data.

\IGNORE{

Re triton -- say one of the keys to success for triton is that it asks the user to explicitly talk about the partitioning of the computation into blocks. The fact that the programmer does that explicitly means that the compiler doesn't have to figure that out with heuristics. The challenge is that you can only do it once (rather than for the other layers). We're applying that pattern repeatedly, and talking about the data. Triton doesnt have to rely on a magic compiler, we dont have to rely on a magic compiler at any level. In principle a compiler could do it, but we dont have one. Compiler can pick where to put things mainly, but partitioning is kind of an algorithm concern.

}
\section{Compilation}\label{sec:compilation}

\begin{figure}
\centering
\begin{minipage}{.43\textwidth}
\centering
\includegraphics[width=0.5\textwidth]{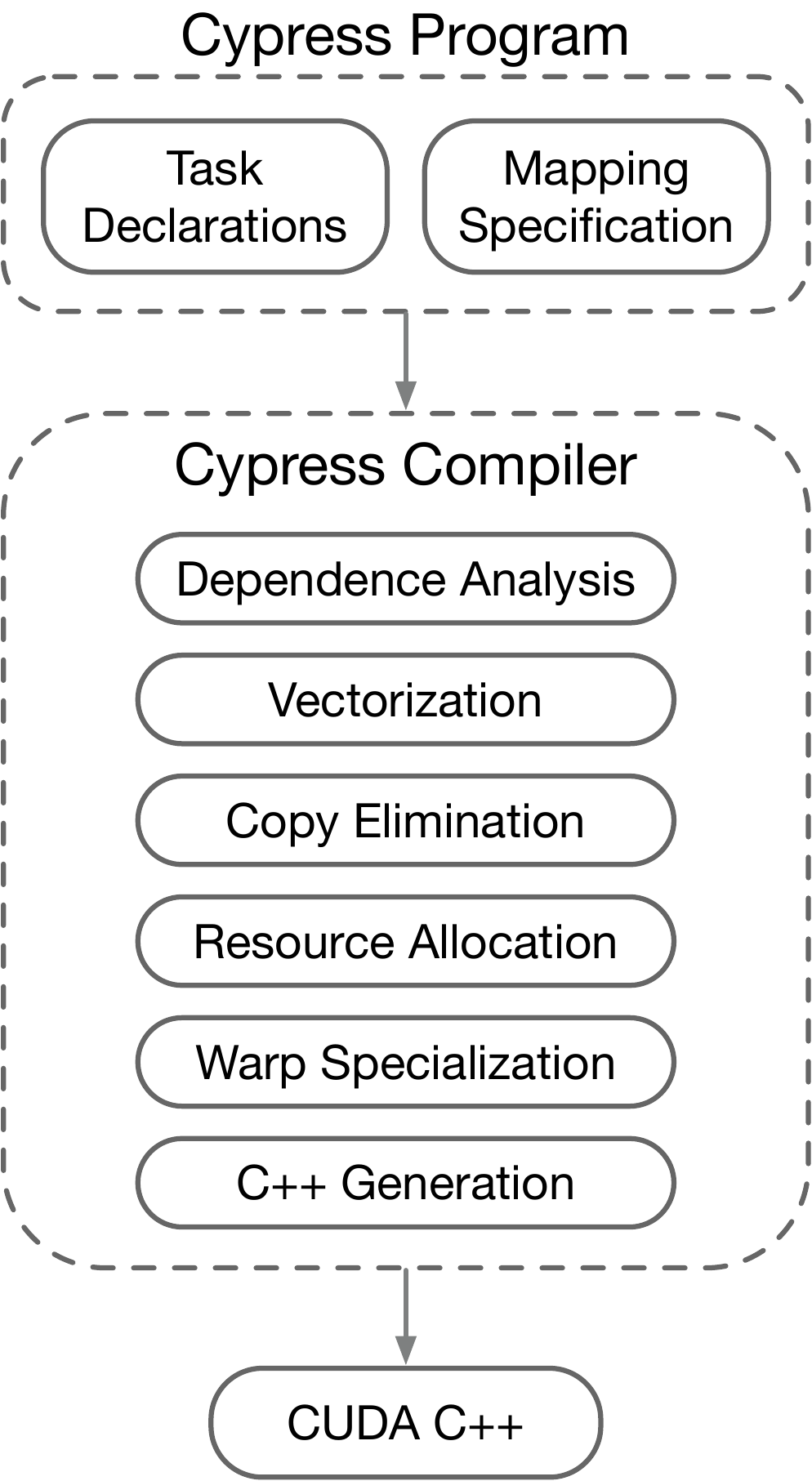}
\end{minipage}\hfill%
\begin{minipage}{.53\textwidth}
\centering
\scriptsize
\[
\begin{array}{rlcl} 
    \textsf{Processor} & p & & \\
    \textsf{Memory} & m & & \\
    \textsf{Variable} & x & & \\
    \textsf{Tensor} & t & \bnfdef & (\textsf{int list}, m) \\
    \\

    \textsf{EventType} & et & \bnfdef & () \bnfalt (\mathbb{N}, p)~\textsf{list} \\
    \textsf{EventIndex} & ei & \bnfdef & e \bnfalt~: \\
    \textsf{Event} & ev & \bnfdef & x \bnfalt ev[\overline{ei}] \\
    \textsf{Precondition} & pr & \bnfdef & ev~\textsf{set} \\
    
    \\

    \textsf{Expression} & e & \bnfdef & x \bnfalt t \bnfalt \mathbb{Z} \bnfalt e \oplus e \bnfalt \\
    &&& \textsf{partition}(\overline{e}) \bnfalt e[\overline{e}] \\ 


    \textsf{Operation} & o & \bnfdef & x = e \bnfalt \\
    &&& ev : et = \textsf{copy}(e, e), pr \bnfalt \\
    &&& ev : et = \textsf{call}(f, \overline{e}), pr \bnfalt \\
    &&& ev : et = \textsf{for}~x~\textsf{in}~[e, e), pr ~\textsf{do}~b \bnfalt \\
    &&& ev : et = \textsf{pfor}~x~\textsf{in}~[e, e), pr ~\textsf{do}~b \\

    \textsf{Block} & b & \bnfdef & \overline{o}; \textsf{yield}~ev \\
    
\end{array}
\]
\end{minipage}

\begin{minipage}[t]{.43\textwidth}
    \captionof{figure}{\name{} compiler architecture overview.}
    \label{fig:compiler-overview}
\end{minipage}\hfill%
\begin{minipage}[t]{.53\textwidth}
    \caption{\name{}'s event-based intermediate representation.}
\label{fig:ir-grammar}
\end{minipage}

\end{figure}

The Cypress compiler lowers a logical program description and a mapping specification to warp-specialized CUDA C++ with explicit communication and synchronization.
We chose to generate CUDA C++ as a simplifying decision for the prototype
compiler; a lower-level target like NVVM could be used in a production implementation.
At the fine-grained scale of code running in the SM, there is no room for
the overhead of a dynamic runtime system that schedules tasks and copies.
Instead, the Cypress compiler leverages and then eliminates all task-based abstractions
to generate statically scheduled loops.
The architecture of the Cypress compiler is shown in \Cref{fig:compiler-overview}.
We discuss the intermediate representation (IR) used by the Cypress compiler and 
then discuss each of
the passes involved in lowering
 \name{} programs to CUDA C++.


\subsection{Intermediate Representation}\label{sec:ir}

A grammar for a simplified form of \name{}'s intermediate representation is shown in
\Cref{fig:ir-grammar}.
The IR contains operations like explicit copies between tensors and task invocations, 
as well as sequential and parallel for loops.
Any potentially asynchronous operation in the IR (such as a copy or an invocation of a leaf task via the \textsf{call} operation)
produces an \emph{event} that represents the completion of the operation. 
Each asynchronous operation accepts a set of precondition events 
that must complete before it can start.
Consequently, the IR encodes a dependence graph through the event connections between operations.
%
%
The \textsf{for} and \textsf{pfor} operators have a \texttt{Block}
which is the loop body; this block has its own completion event to signal the end of an iteration.
Scalar computations, such as index arithmetic or creating and 
accessing a partition, use a standard representation 
without events.
%
While not shown, the IR contains standard control-flow constructs
like branches, which are adapted to yield events in each divergent path.
%
The IR is in a single-static-assignment (SSA) form to ensure that any valid ordering of operations forms an ordering that satisfies all event dependencies.

The most interesting component of \name{}'s IR is the representation of events.
An event type is either unit or an array of events, where each dimension in the
array is annotated with a processor kind.
Event arrays are created by parallel loops, where each element of the array 
corresponds to the completion event of an iteration on a particular processor.
An event array can be indexed to extract the particular event that
a future operation should depend on.
Indexing can be done with an integer value or the broadcast 
operator \texttt{[:]} (inspired by NumPy broadcasting~\cite{numpy}),
which returns an event representing all events along the indexed dimension completing.
Indexing an event array represents synchronizing processors of the indexed dimension.
We leverage this representation of events in \Cref{sec:analysis:vectorization}
and \Cref{sec:analysis:copy-elim} to track synchronization of parallel loop
iterations.
We emphasize that events in \name{}'s IR are an intermediate
construct only, and are not realized in generated code; there is no dynamic dependence tracking performed at run time.

\subsection{Compiler Architecture}

The Cypress compiler consumes a program in the form of a logical description and mapping specification and transforms it into a CUDA C++ program.  
The \name{} compiler performs this transformation through a series of passes over the IR, shown in \Cref{fig:compiler-overview}.
The first three passes (dependence analysis, vectorization and copy elimination) capture important information from the task-based representation of the program and then lower away the tasking abstractions,
while the next two (resource allocation and warp specialization) perform optimizations.
The final stage (lowering to C++) replaces events by system-specific synchronization   and performs other
mechanical translations required to construct valid CUDA programs.
We discuss each pass in turn.

\subsubsection{Dependence Analysis}\label{sec:analysis:dep-analysis}

\input{dep-analysis}

The first stage  is dependence analysis, which consumes the task-based logical description and the mapping specification (\Cref{fig:frontend-grammar}) to perform a syntax-directed translation to the Cypress IR (\Cref{sec:ir}).
%
%
The dependence analysis inserts event dependencies to enforce ordering between tasks using the same data with true or anti-dependencies~\cite{dragon-book}, while tasks operating on independent data or with non-interfering privileges (e.g., both reading) may execute in parallel.
Additionally, the analysis must also insert data movement to maintain
coherence when the same logical tensor is mapped to different memories by distinct tasks.
Once the dependence analysis has inserted all dependencies required to enforce
the sequential semantics of the source program, later passes must preserve
these dependencies throughout applied transformations.

The dependence analysis is an in-order traversal of the instantiated task tree, starting at the task variant denoted as the entrypoint
of the computation in the mapping specification.
The traversal maintains an event for each tensor in the context of the task variant.
At each task launch site, the privilege annotations of the
launched task are used to register the appropriate events as preconditions for the task and to update
the set of events with the completion event of the task launch.
For example, if a task writes to a tensor, then later readers of the tensor 
record an event dependence on the writing task's completion.
Dependencies are enforced 
between task invocations through chaining events in the IR.
To lower a task launch, 
the compiler consults the mapping specification 
to determine the invoked task variant and the memory where each
tensor argument should be placed.
For all tensor arguments to the callee task, the dependence analysis employs a \emph{copy-in/copy-out} discipline. 
Lowering a task launch into the Cypress IR consists of four steps:
\begin{enumerate}
    \item Create a fresh allocation for each tensor argument to the callee task, in the memory specified by the mapping,
    \item For each tensor argument read by the callee task, emit a copy from the existing tensor allocation into the fresh allocation and record any event preconditions for the copy,
    \item Record all copy completion events as preconditions for the callee task and then recursively traverse the selected task variant of the callee to generate its IR, and
    \item For each tensor argument written by the callee task, emit a copy from the fresh allocation into the existing allocation of the caller task with the callee completion event as a precondition.
\end{enumerate}
When lowering a parallel group of tasks, the broadcast indexing
operator is used to condition future operations on the completion of all parallel iterations.
The copy-in/copy-out discipline ensures dependence analysis is always local
to the task variant being traversed,
which minimizes the complexity of the compiler.
%
%
%
%
This approach does introduce some unnecessary copies, but a robust copy elimination pass (\Cref{sec:analysis:copy-elim}) ensures that such copies are removed.

%


An example of the dependence analysis on a subset of the GEMM program (\Cref{fig:cypress-gemm})
is in \Cref{fig:dep-analysis-example}, which includes a portion
of the previously omitted \texttt{clear} task tree.
For simplicity, we consider a version 
that does not target the warpgroup level, so
tasks running at the block level launch sub-tasks targeting the warp level.
The example starts the dependence analysis procedure to lower the
\texttt{gemm\_block} task variant.
%
\texttt{gemm\_block} launches the
\texttt{clear} task, which we assume is mapped to the \texttt{clear\_inner} variant.
The analysis creates a fresh tensor (\texttt{C1}) for the argument,
and uses \texttt{C1} for the recursive lowering of \texttt{clear\_inner}.
The \texttt{prange} is lowered into a \texttt{pfor} loop with
a fresh allocation (\texttt{CW}) for the recursive launch of \texttt{clear} (\Cref{fig:dep-analysis-example-post}, Line~\ref{fig:dep-analysis-clearpfor1}).
%
\texttt{clear\_inner} is instantiated again for the \texttt{THREAD}
processor level, so \texttt{CW} is partitioned
and another \texttt{pfor} with a fresh allocation \texttt{CR}
is emitted (\Cref{fig:dep-analysis-example-post}, Line~\ref{fig:dep-analysis-clearpfor2}).
As \texttt{clear\_inner} writes to \texttt{CR}, 
a copy-out is emitted from \texttt{CR} to the corresponding slice of \texttt{CW} (\Cref{fig:dep-analysis-example-post}, Line~\ref{fig:dep-analysis-copyout1}).
The loop body yields the completion event of the final operation in the loop.
%
A similar copy-out is emitted from \texttt{CW} to \texttt{C1}, but this copy depends on \texttt{e2[:]}, indicating that all
threads must complete their iteration before the copy starts, as each
thread writes to a slice of \texttt{CW} (\Cref{fig:dep-analysis-example-post}, Line~\ref{fig:dep-analysis-copyout2}).
Finally, a copy-out for the inline launch of \texttt{clear} in \texttt{gemm\_block}
is emitted between \texttt{C1} and \texttt{C}, also with a broadcasted precondition.
The resulting event \texttt{e6} is registered as the ``current'' event
for readers of \texttt{C}.

Next, the \texttt{srange} over \texttt{gemm} tasks is lowered.
The emitted \texttt{for} loop depends on \texttt{e6}, since
the \texttt{gemm} task reads from \texttt{C}.
Within the body of the loop, the analysis creates 
fresh tensors for \texttt{gemm}'s arguments, and issues
copy-ins for each of the tensors.
After the similarly elided recursive lowering of \texttt{gemm}, 
a copy-out for the only written-to tensor (\texttt{C2}) is performed.
We elide the resulting IR for lowering the launch of the
final \texttt{copy} task, which is similar to  the \texttt{clear} task.
The result of the dependence analysis is a
graph of asynchronous tasks and copies linked by events.

\subsubsection{Vectorization}\label{sec:analysis:vectorization}

Next, \name{} performs a vectorization process that flattens the nested loop
structure of programs created by the dependence analysis.
This pass removes nested loops that are implicit in the GPU programming model,
such as \texttt{pfor} loops over the warpgroups, warps and threads.
The vectorization process leverages the IR's indexable event arrays
to preserve the dependencies between parallel iterations after the parallel
loops are flattened.
The mechanics of the vectorization are straightforward: starting from
the deepest nesting, each implicit parallel loop is flattened, and the iteration
variable is substituted with an expression that evaluates to the processor index (such
as the warp or thread index).
All event arrays created within a flattened implicit loop are \emph{promoted}
by adding an additional dimension with size equal to the extent of the flattened loop.
Then, any consumers of events within the implicit loop are rewritten
to index each event array with the processor index.
Point-wise dependencies are preserved between the independent iterations of
a parallel loop, while synchronization needed before copies and
dependent tasks are encoded with broadcasted indexing of events.
The vectorization process is shown in 
\Cref{fig:vectorization}, where the loops over the warps and threads
in the program from \Cref{fig:dep-analysis-example} are flattened away.
Both the point-wise dependencies between the parallel 
iterations (\Cref{fig:vec-final}, Line~\ref{fig:vec-pointwise-event})
and the post loop synchronizations are preserved (\Cref{fig:vec-final}, Line~\ref{fig:vec-bcast-event}).

\input{vectorization}

\subsubsection{Copy Elimination}\label{sec:analysis:copy-elim}

The third stage is a copy elimination pass, which is critical
to remove unnecessary copies introduced by
the copy-in/copy-out discipline in the dependence analysis (\Cref{sec:analysis:dep-analysis}).
The pass consists of a set of rewrite patterns that remove
or move copies, which are similar to patterns in the Sequoia compiler~\cite{sequoia-compiler}.
A subset of patterns is shown in \Cref{fig:copy-elim-patterns}.
These patterns include straightforward optimizations like eliminating copies
between the same tensor allocation and duplicate copies into the same tensor allocation.
The spill elimination and spill hoisting patterns leverage the structure
of the machine and data model.
%
The spill elimination pattern (\Cref{fig:spill-elimination}) erases copies where a 
tensor is copied into a slice of its parent, and then the parent slice is copied
back into the tensor.
The spill hoisting pattern (\Cref{fig:spill-hoisting}) identifies when a copy is performed from a parent
tensor into a child tensor, and then back from the child into the parent within a loop.
These copies can be hoisted into the preamble and postamble of the loop.
%

The pass leverages the structure of eliminated copies to elide or preserve
synchronization.
Consider \Cref{fig:spill-elimination} in which two parallel blocks each use the same partition and there is a copy-out followed by a copy-in using the same partition between the blocks.
The second copy collapses the event array \texttt{e2} prior to copying from the temporary allocation \texttt{t} because the dependence analysis needed to ensure that all the data in \texttt{t} was ready before any copy-in operations could begin. 
In general this is necessary for correctness, especially in the cases where the copy-out and copy-in operations could be using different partitions.
However, in the case where the same partition is used, then we can safely elide both the copies as well as the synchronization implied by the collapsing of the \texttt{e2} event array between the copies.
This recovers the expected behavior that there are only point-wise dependencies between the iterations of \texttt{b1} and \texttt{b2} after the copies are removed.
%
%
%
%

\input{copy-elim}

In contrast, copies removed by patterns like the self copy elimination pattern (\Cref{fig:self-copy-elim})
forward their event array dependencies directly.
%
In these patterns, any cases where event arrays are collapsed cannot be removed as
the resulting tensor is valid only when all parallel copies complete.
An example of this case is when an entire warpgroup copies the accumulator tensor from
the register memory into the shared memory for the TMA to consume.
Copies between the thread-local and warp-local partitions of the shared memory tensor
are removed, but the synchronization of the warpgroup must remain before
notifying the TMA that the shared memory buffer is ready to be consumed.

In order to eliminate as much synchronization as possible during the copy
elimination pass, the order that rewrite patterns is applied can be important.
In particular, we order patterns that can eliminate events (spill-related patterns) ahead
of patterns that must preserve existing dependencies.
While not currently supported in MLIR, utilizing an equality saturation framework like an
e-graph~\cite{egg} could avoid the need for these ordering heuristics.

\subsubsection{Resource Allocation}\label{sec:analysis:resource-alloc}

After the copy elimination pass removes copies and intermediate tensors, 
remaining tensors must be bound to physical allocations within the memory 
they have been mapped to.
We focus on the allocation of tensors within each SM's shared memory as it is often the most constrained memory on-chip.
While other compilers~\cite{triton, sequoia-compiler} have studied
statically allocating tensors in shared memory, the asynchronous
environment requires a different strategy.

The main decision to make when allocating within a fixed size memory is how to trade-off memory pressure with parallelism. 
For Cypress this trade-off requires deciding which logical tensors should be placed in the same or overlapping physical addresses at different times---i.e., which tensors should be aliased onto the same physical memory.
%
%
Some aliasing of logical tensors is nearly always required as the minimum size of tensors required by the Tensor Cores is a significant fraction of shared memory, which limits the number of logical tensors that can be live at a time.
Additionally, better performance can arise from executing multiple thread blocks simultaneously on an SM, requiring a partitioning of the shared memory that further limits the capacity available to a single thread block and mandates more aliasing.
However, too much aliasing decreases the degree of parallelism, 
exposing latency and degrading performance. 
%
\name{}'s approach has users specify an upper bound 
on shared memory usage for each thread block in the mapping specification; 
the compiler then allocates tensors with as little aliasing
as possible to maximize parallelism.
This policy allows users to make the footprint-occupancy trade-off for thread blocks while maximizing performance within the given constraints. 

\begin{figure}
\centering
\begin{minipage}{.48\textwidth}
    \centering
    \includegraphics[width=0.8\linewidth]{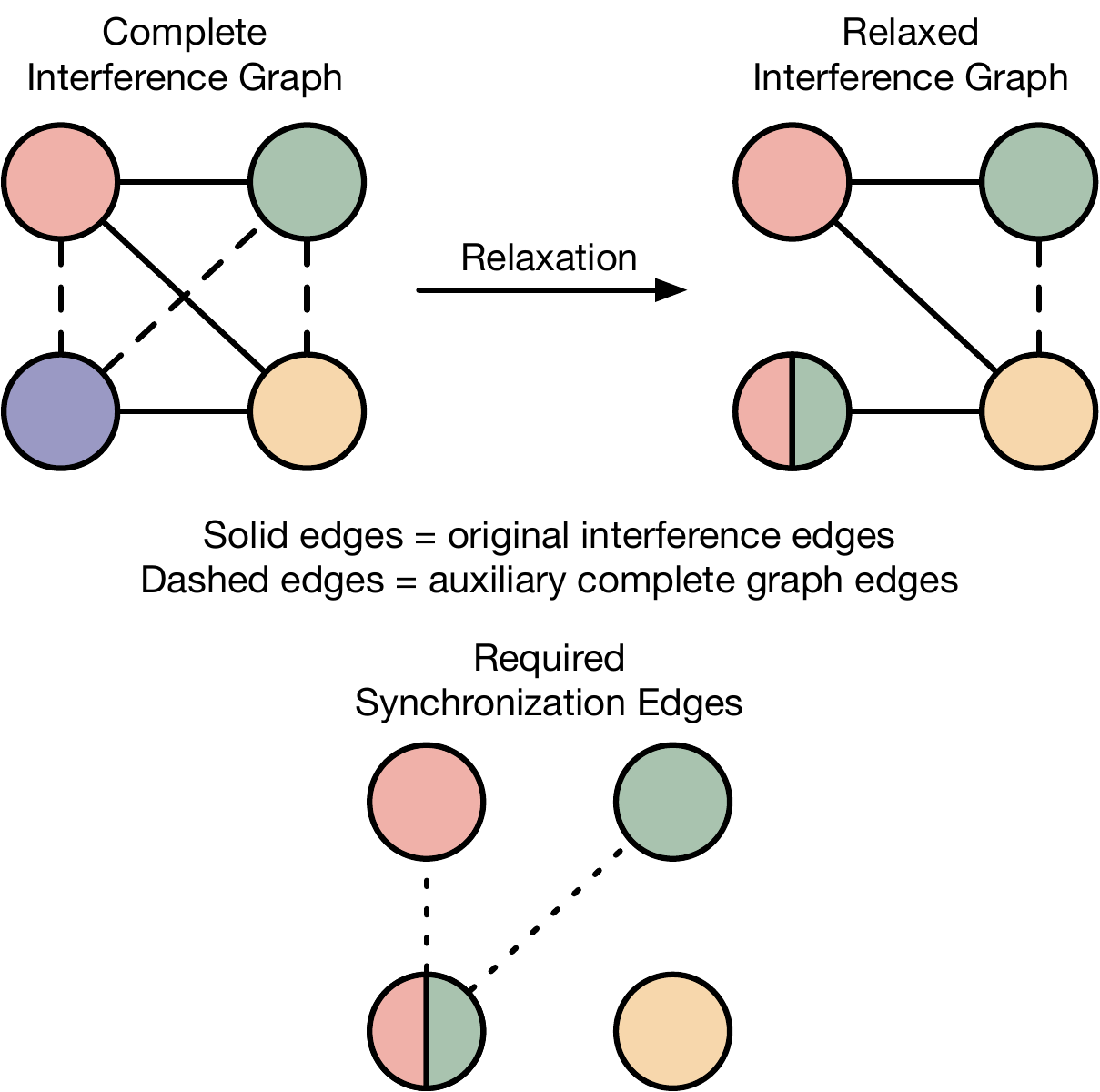}
\end{minipage}\hfill%
\begin{minipage}{.48\textwidth}
    \centering
    \includegraphics[width=0.8\textwidth]{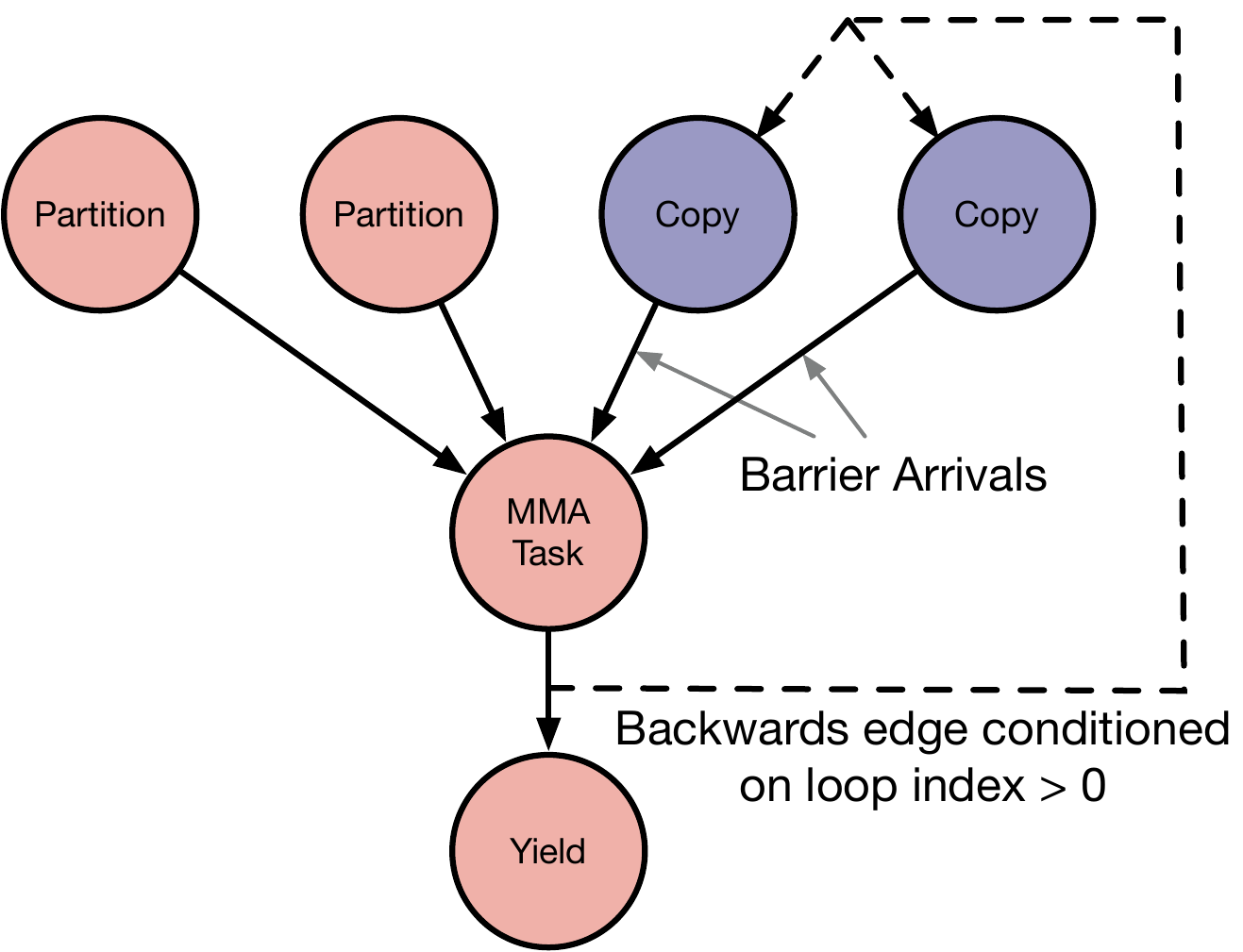}
\end{minipage}

\begin{minipage}[t]{.48\textwidth}
    \captionof{figure}{Resource allocation in \name{}. Nodes and colors indicate tensors and
    allocated buffers.}
    \label{fig:resource-alloc}
\end{minipage}\hfill%
\begin{minipage}[t]{.48\textwidth}
    \captionof{figure}{Warp specialization of the GEMM main loop dependence graph. Colors indicate warp assignment.}
    \label{fig:warp-specialization}
\end{minipage}

\end{figure}

%
\name{} constructs an interference graph out of all shared memory tensors,
and then adds \emph{auxiliary} edges to complete the interference graph to 
force all tensors into independent allocations.
Cypress then iteratively attempts to construct an allocation assignment that fits within the user
provided limit using the strategy described by Knight et al.~\cite{sequoia-compiler}.
Starting with the complete interference graph, Cypress removes auxiliary edges from 
the interference graph until an allocation assignment is feasible.
If allocation is impossible on the original interference graph, then an out-of-memory
error is reported to the user, who must adjust their mapping specification to place fewer tensors in shared memory or increase the shared memory for each thread block.
By starting with the complete interference graph and removing edges, Cypress ensures that a selected assignment performs a minimal amount of aliasing. 
This allocation strategy is visualized in \Cref{fig:resource-alloc}.


Once an allocation strategy has been found, the compiler must insert additional dependencies to ensure that the live ranges of logical tensors assigned to the same physical allocation do not overlap.
%
%
%
Starting from the relaxed interference graph that resulted in a successful allocation strategy, 
\name{} identifies the necessary dependencies to insert by collecting
all tensors that
do not share an interference edge.
Since these tensors will be assigned to the same allocation, Cypress inserts event dependence edges between the last readers of one logical tensor and the first writer of the next logical tensor using the allocation to avoid write-after-read hazards.
%

\subsubsection{Warp Specialization}\label{sec:analysis:warp-specialization}

Warp specialization~\cite{singe-compiler, cuda-dma} 
is a GPU programming technique that
partitions the computation across multiple warps within a thread block,
exposing concurrency between the warps and allowing for resources to be 
distributed across the warps.
%
Warp specialization has become an effective technique to target
the Hopper architecture, at the cost of increased software complexity.
%
When warp specialization is requested by a mapping specification, Cypress 
separates the computation into multiple compute warps and a data movement (DMA) warp 
to 
prevent interference when interacting with different fixed function units (e.g., the TMA and Tensor Core),
and to allow for a larger fraction of the register file to be used by the compute warps.
%
Cypress treats warp specialization as a graph partitioning problem~\cite{singe-compiler}, where
the dependence graph encoded in the IR must be partitioned between the DMA and compute warps.
The compiler constructs a partition where all global-to-shared copies are assigned
to the DMA warp, and all other operations are assigned to the
compute warps.
Additional warp specialization strategies could be implemented in the \name{} compiler
by changing the chosen partitions of the dependence graph.
Any dependence edges that cross between the two partitions of the graph 
indicate synchronization that must be performed between the warps using barriers.
%
%
Cypress handles the warp specialization of loops separately, both to manage
correctness and to perform additional optimizations.
An example of the graph partitioning and crossing dependence edges is shown in \Cref{fig:warp-specialization}.


To optimize warp-specialized loops, the compiler supports a
\emph{pipelining} transformation to pre-fetch data and avoid exposed global memory access latency.
\name{} pipelines a loop's dependence graph by
unrolling it to the pipeline depth specified by the mapping, and then compacting the loop back
into a single iteration.
Pipelined tensors and events in the compacted body are indexed by the loop variable
modulo the pipeline depth.
The desired result of pipelining can be seen in \Cref{fig:hopper-gemm}, where
the shared memory tiles of $A$ and $B$ have an extra dimension with size equal to 
the pipeline depth \texttt{PIPE}, and are correspondingly indexed in their uses.
When using warp-specialization, the data movement warp runs \texttt{PIPE} iterations
ahead of the compute warps, pre-fetching data with the TMA to hide global memory access latency.
We found this pipelining transformation to generate better code than
explicit loop unrolling due to lower register pressure and instruction
cache footprint.

When pipelining, \emph{backwards dependencies}
are inserted into the dependence graph to enforce write-after-read anti-dependencies.
In particular, asynchronous copies with no preconditions, such as those that would be inserted for
tiles of $A$ and $B$ inside the main $K$-reduction loop of matrix multiplication, must only begin
once the consumers of the copies' destination buffers from prior iterations have completed.
These backwards dependencies can be seen in the hand-written GEMM sketch in \Cref{fig:hopper-gemm},
and are denoted with a dashed line in \Cref{fig:warp-specialization}.
%
%
%

\subsubsection{CUDA C++ Generation}\label{sec:analysis:c++lowering}

%
The code generation phase handles the low-level details of generating valid
CUDA, such as outlining device functions and launching kernels.
%
%
The most important component is lowering events in the IR onto 
hardware-specific synchronization primitives on the GPU.
Events produced by copies and tasks are lowered to the corresponding
synchronization mechanism of the producing operation.
For example, events produced by TMA copies are lowered into shared memory barrier
arrivals that the TMA triggers, and events produced application tasks using the Tensor Core
are lowered onto the Tensor Core synchronization assembly instruction.
%
%
Events that cross warp boundaries as a result of warp specialization are lowered onto shared
memory barriers.
\footnote{While prior work on warp specialization used named barriers for
synchronization~\cite{cuda-dma,singe-compiler}, shared memory barriers are required to synchronize warps across multiple CTAs when
the Hopper TMA multi-casting feature is used.}
Event arrays are lowered by specializing on the array's indexing pattern.
Event arrays indexed only with processor indexes can be removed, as the
implied point-wise dependence is satisfied by a valid SSA ordering of the IR operations.
Event arrays indexed in a broadcasted manner are lowered into
different synchronization primitives depending on the processor annotation
of the broadcasted dimensions.
A broadcast at the thread level is lowered to a \texttt{\_\_syncwarp} function, while a broadcast at the warp or warpgroup level is lowered to a named barrier arrival and wait~\cite{cuda-dma}.
%

\section{Evaluation}\label{sec:evaluation}

%
We evaluate \name{} on a variety of compute-limited linear algebra kernels
that utilize the TMA and Tensor Core on Hopper.
We show that kernels developed using \name{} can achieve performance competitive
to expert-written kernels in \cublas{} and \cudnn{}, as well as reference kernels
developed by experts using CUTLASS~\cite{cutlass} and ThunderKittens~\cite{thunderkittens}.
These expert-written kernels contain explicit data movement, synchronization
and warp specialization in order to achieve high performance.
We also show that when compared to a higher-level language like Triton~\cite{triton},
the mapping control and first-class asynchrony of \name{} enables programs
to achieve higher performance.
Our results show that the \name{} compiler is able to generate high performance GPU computations
while automating data movement and synchronization for the programmer.

\subsection{Experimental Setup}
We evaluate \name{} on an NVIDIA H100 80GB SXM5 GPU.
We used CUDA 12.5.1 for most experiments except for the Flash Attention experiments, 
where we found that different systems (including \name{}) were sensitive to
the version of CUDA.
For each system, we chose the CUDA version that delivered the best performance.
Flash Attention 3~\cite{flash-attention-3} and cuDNN achieved the best performance
on CUDA 12.3.1.
We benchmarked \name{}'s Flash Attention 3 with the most recent build
of the NVCC compiler.
%
%
We compare against Triton nightly \texttt{3.0.0.post20240716052845},
and directly use (or adapt) the publicly available Triton example programs
for our benchmarks.
All results are the average of 100 iterations with 5 warmup iterations.
For GEMM-like computations that approach the peak throughput of the
device, we benchmark with values drawn from the same random distribution of matrix elements across systems
to normalize the effects of power throttling, which we observed to have a significant performance impact.
The mapping for each \name{} program were developed by us and manually tuned; the individual
mapping strategies chosen were informed by existing algorithms and implementations.

\subsection{GEMM Kernel Variants}

%
%


\paragraph{GEMM and Batched-GEMM}

\begin{figure}
\begin{subfigure}[h]{0.24\textwidth}
    \centering
    \includegraphics[width=\textwidth]{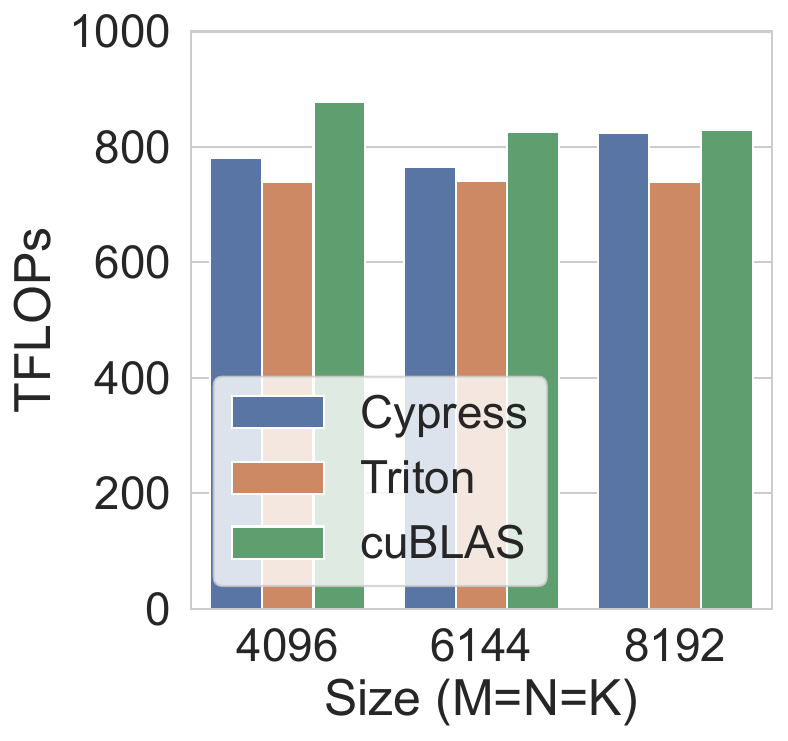}
    \caption{GEMM}
    \label{fig:gemm}
\end{subfigure}\hfill%
\begin{subfigure}[h]{0.24\textwidth}
    \centering
    \includegraphics[width=\textwidth]{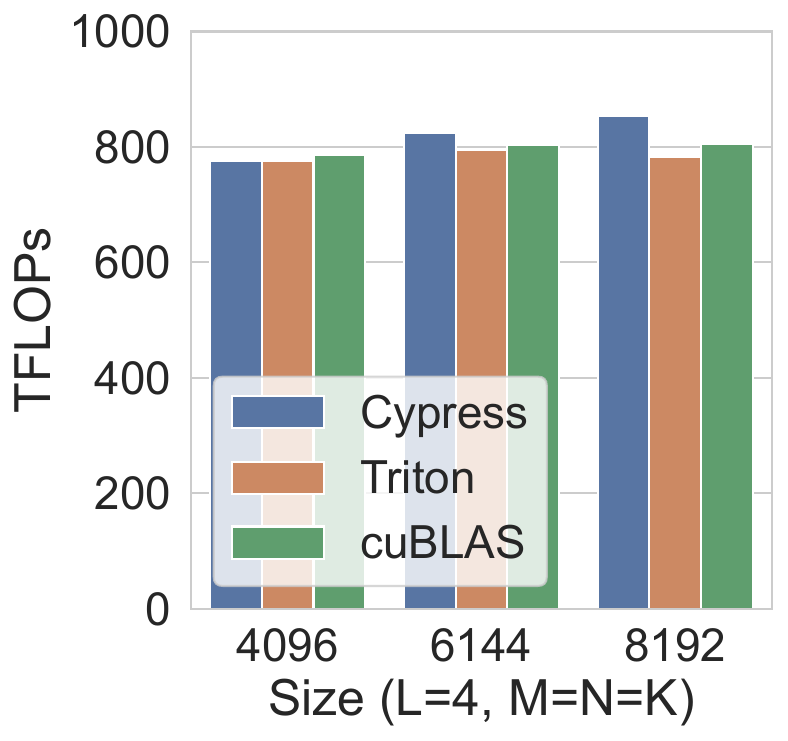}
    \caption{Batched-GEMM}
    \label{fig:batched-gemm}
\end{subfigure}\hfill%
\begin{subfigure}[h]{0.24\textwidth}
    \centering
    \includegraphics[width=\textwidth]{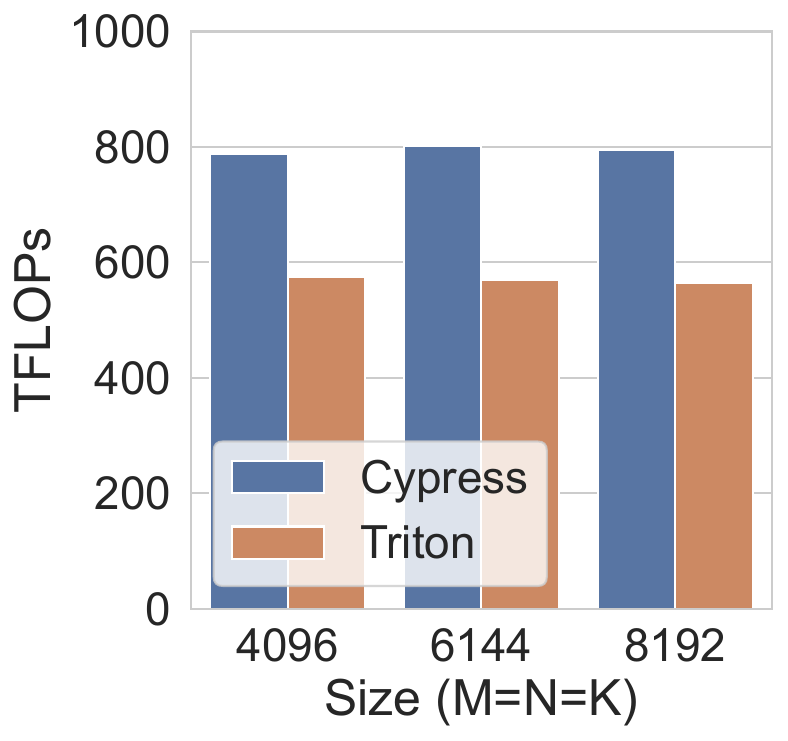}
    \caption{Dual-GEMM}
    \label{fig:dual-gemm}
\end{subfigure}\hfill%
\begin{subfigure}[h]{0.24\textwidth}
    \centering
    \includegraphics[width=\textwidth]{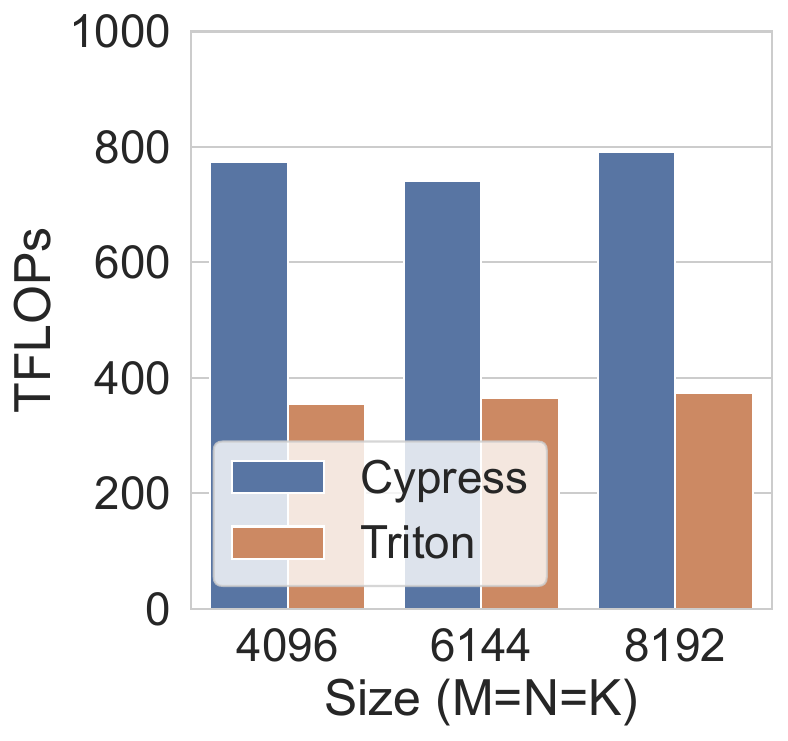}
    \caption{GEMM+Reduction}
    \label{fig:gemm-reduc}
\end{subfigure}
\caption{Throughput on FP16 GEMM Kernel Variants.}
\end{figure}

We first evaluate \name{} on a standard FP16 GEMM (as seen in \Cref{fig:cypress-gemm}), and compare with
cuBLAS and Triton.
\Cref{fig:gemm} shows that \name{} can generate code that achieves 0.88x-1.06x the performance of \cublas{}
and slightly outperforms Triton (1.05-1.11x).
\name{}'s implementation hierarchically decomposes the GEMM within each thread block 
across multiple warpgroups to fit larger tiles in shared memory and fill the Tensor Core
from independent pipelines.
We also evaluate \name{} on a Batched-GEMM program, where $L$ independent GEMMs
are performed in a single pass, shown in \Cref{fig:batched-gemm}.
Similarly to the standard GEMM computation, \name{} performs competitively with
cuBLAS and Triton, even slightly outperforming cuBLAS on the largest problem size.
GEMM and Batched-GEMM are important and heavily optimized computations, and \name{}
is able to generate code that is competitive with both hand-tuned and popular compiled implementations.

\paragraph{Dual-GEMM}
We now turn to fused computations that are not part of standard BLAS libraries.
We consider Dual-GEMM, which computes $A \cdot B_1 + A \cdot B_2$
in a single kernel, avoiding storing temporaries in
global memory.
Dual-GEMM is a core computation in Gated Linear Units~\cite{glu-shazeer, dauphin-glu}, a neural network
layer used in Transformer networks.
A Dual-GEMM should achieve similar performance to a GEMM by
overlapping the independent GEMMs within the main loop and overlapping
the asynchronous copies of $B_1$ and $B_2$.
\name{} performs these optimizations automatically, only inserting synchronization
when required to maintain sequential semantics, and achieves similar throughput
as GEMM (\Cref{fig:dual-gemm}).
In contrast, Triton experiences a degradation in performance
issuing an additional GEMM inside the main loop.
\name{} achieves 1.36x-1.40x the performance of Triton.
We investigated the generated Triton IR and saw that Triton does not partially
overlap loads of $B_2$ while computing $A \cdot B_1$.
Also, Triton does not use the TMA by default, requiring modification of the
kernel and launching code to explicitly use an experimental TMA operation.

\paragraph{GEMM+Reduction}
Next, we consider a fused GEMM+Reduction kernel that computes
$C = A \cdot B$ and $y(i) = \sum_k A(i, k)$ in a single kernel.
The reduction of $A$ can be overlapped on the SIMT threads while the
Tensor Core is asynchronously computing $A \cdot B$.
\name{} exploits asynchrony to perform
the GEMM+Reduction kernel at a similar throughput to a standard GEMM, achieving a 2.02-2.18x speedup over Triton (\Cref{fig:gemm-reduc}).
%
We find that Triton does not overlap the GEMM computation with
the reduction, explicitly waiting on the Tensor Core before issuing
the reduction.
Second, we find that Triton heuristically places the reduction accumulator
in the shared memory of the SM, while our implementation with \name{} uses the
mapping specification to place the accumulator inside the register file.
We were able to adjust the \name{} mapping and modify the \name{} generated
code to reproduce the performance achieved by Triton.
This benchmark highlights the importance of first-class asynchrony and control over mapping decisions in \name{}.

\subsection{Flash Attention}

\begin{figure}
    \centering
    \includegraphics[width=0.85\linewidth]{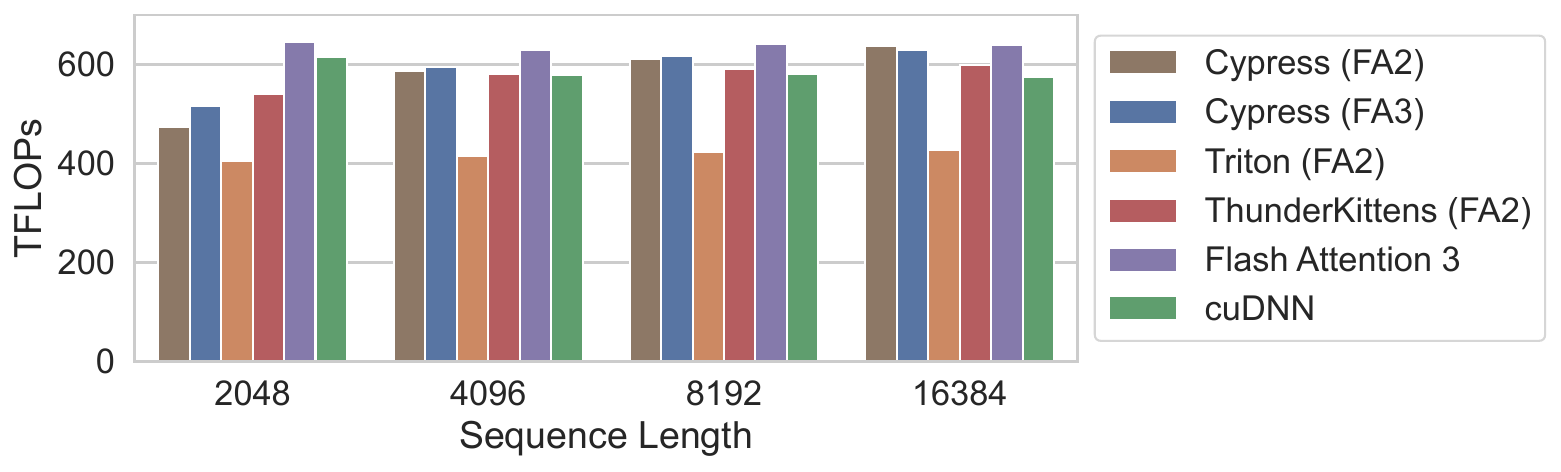}
    \caption{FP16 Flash Attention Throughput (HeadDim = 128).}
    \label{fig:flash-attention}
\end{figure}

The most complex \name{} programs we have developed are variants of the
forward attention kernel widely used in transformer models.
We implement the Flash Attention 2~\cite{fa2} and Flash Attention 3~\cite{flash-attention-3} algorithms.
%
We compare against several expert-written reference implementations: Triton, cuDNN, ThunderKittens~\cite{thunderkittens}
and the original Flash Attention 3 implementation and show that
\name{} is competitive with the best-known implementations (\Cref{fig:flash-attention}).

An implementation of Flash Attention 2 that targets Hopper must leverage the
TMA and Tensor Cores available on Hopper to overlap communication and
computation when possible to achieve reasonable performance~\cite{fa2-hopper}.
\name{} assists with these goals, automatically leveraging the TMA and performing
overlap of the matrix multiplications with reduction initialization when possible.
Unlike the Flash Attention 2 implementation using ThunderKittens~\cite{thunderkittens},
the \name{} program abstracts implementation details such as warp specialization, asynchronous copies, barriers and
Tensor Core synchronizations, while still delivering comparable performance (between 0.87x-1.06x).
We discuss the Flash Attention 2 performance in more depth after discussing Flash Attention 3.

Flash Attention 3~\cite{flash-attention-3} extracts more parallelism by 
restructuring Flash Attention 2.
In particular, the Flash Attention 2 main loop contains (at a high level) a GEMM, whose results
are used in a row-wise reduction, which is then used in a second GEMM.
This sequential dependency limits the amount of work that can be done in parallel with matrix
multiplications on the Tensor Core.
To improve performance, Flash Attention 3 creates a copy of the
results of the first GEMM, and performs a manual software-pipelining transformation
to allow for the reduction of iteration $k$ to be overlapped with the first
GEMM of iteration $k+1$.
%
To overlap operations correctly, the algorithm description in Flash Attention 3
is presented with specific locations that communication and synchronization of the
TMA and Tensor Core should be placed.
When implemented in \name{}, once the main loop has been rewritten in the desired
pipelined manner, the \name{} compiler \emph{infers all} of the interleaved
communication and synchronization described manually in the Flash Attention 3 work.
The \name{} implementation of Flash Attention 3 performs competitively with the
reference Flash Attention 3 implementation, achieving between 0.80x-0.98x
the performance.

\name{}'s Flash Attention implementations are competitive with hand-tuned
implementations (ThunderKittens, Flash Attention 3 and cuDNN), and outperform the high-level
Triton implementation.
Control over mapping and partitioning allows \name{} to implement 
the strategies used in hand-tuned implementations, while automating communication
and synchronization like Triton.
An interesting result was the similar performance between Flash Attention 2
and Flash Attention 3.
We used a similar set of tuning parameters as ThunderKittens, which differ
from prior published Flash Attention 2 implementations for Hopper~\cite{fa2-hopper}.
In particular, we adjusted the mapping specification to use three
consumer warpgroups rather than two, and let the warp scheduler
interleave the Tensor Core and SIMT work of the three warpgroups.
We found that with an extra consumer warp group and less register pressure from 
no longer storing a copy of the first GEMM's accumulator, we could achieve
competitive performance with the Flash Attention 3 implementation, indicating the difficulty of discovering
the fastest kernels in a jagged performance landscape.
The performance gap between \name{} and the reference Flash Attention 3 implementation
at small sequence lengths comes from \name{} not yet implementing the \emph{persistent kernel} optimization.
The persistent kernel optimization launches a single CTA per SM on the target
GPU and schedules logical blocks of work onto these persistent CTAs, which
lowers scheduling and initialization overheads for small problems.
Such an optimization could be performed by \name{}, as \texttt{prange}
loops mapped into the \texttt{BLOCK} level of the machine could be lowered
onto a persistent group of CTAs.
%

\subsection{Programming Experience}

Having written a number of interesting programs in Cypress, we now qualitatively 
contrast our experience with developing in peer systems. 
Unlike low-level systems like CUTLASS, \name{} programs do not contain communication
and synchronization, which simplified many aspects of development.
For example, we could easily modify Cypress programs to add functionality without worrying about how the partitioning (and therefore parallelization) of a new block interacted with existing code as the compiler guaranteed sequential semantics.
%
%
%
%
Additionally, extending a \name{} program with additional levels of hierarchy
was possible with minimal code modifications.
For example, our original GEMM implementation used a single warpgroup.
Modification to the final version with multiple warpgroups involved adding new
task variants for the warpgroup level of the processor hierarchy, and modifying
the mapping specification to dispatch to the new variant.

A downside to the \name{} programming model is a tendency to require similar but slightly different boilerplate in the logical description and mapping
specifications.
%
%
In CUTLASS, once data has been partitioned by the programmer, it can
be reused in its partitioned manner easily, but subtle bugs can occur if new code is introduced that assumes a different partition.
In contrast, a \name{} program expresses the desired
partitioning each time a tensor is accessed, which can lead to repetitive
partitioning logic, but guarantees the safety of the program.
%
%
Similarly, the mapping specification can contain redundancy
when multiple task trees are mapped onto the machine the same way.
While the logical description and mapping specification can
be verbose when initially developing the target application,
we have found that the explicit specification of these aspects
of the program makes it easier to modify and communicates the author's intent for future developers.

Finally, we found that the separate mapping specifications in Cypress simplified
aspects of exploring the performance landscape.
Tuning mapping specification parameters to generate kernels with different combinations of task variants, warp specialization, pipeline depth, and placement of data was possible with small modifications to the mapping.
%
While some tuning parameters are expressible with template parameters in CUTLASS, other
parameters like data placement control require non-trivial code changes to explore.
With Triton, exploring the space is not even possible as many of the knobs we would want to manipulate are hard-coded as heuristics in the compiler.

\section{Related Work}\label{sec:related-work}

\paragraph{Task-Based Parallel Programming}
The most similar work to \name{} is Sequoia~\cite{sequoia, sequoia-compiler, sequoia-runtime},
a task-based parallel programming system that targeted machines with deep memory hierarchies.
Sequoia also compiled a task-based program representation onto a hierarchical machine model.
However, the machine model of Sequoia was more restrictive than \name{}, and wouldn't 
have been capable of targeting modern GPUs. 
In particular, Sequoia's machine model was strictly hierarchical: a processor at level $i$
in the hierarchy may only access an attached memory at level $i$, which wouldn't have been capable of describing a modern GPU where multiple levels of the processor hierarchy can access multiple memories. 
Consequently, Sequoia would have struggled to achieve comparable performance to Cypress because it couldn't explore many of the essential mapping strategies.
%
%
%
The DPJ compiler~\cite{dpj} also supported static parallelization of task trees using partitioned data, but did so for shared memory without considering a deep memory hierarchy or GPUs.
There are many other task-based systems~\cite{legion,starpu,dask,ray,parsec,regent} but all of them rely, at least in part, if not entirely, on dynamic runtimes to schedule and execute programs. 
At the instruction level, where we are focused, the overheads associated with any dynamic runtime system are prohibitively expensive.
%
%
%
\name{}'s partitioning system is inspired by the task-based Regent~\cite{regent, deppart}
language.

\paragraph{Hopper Programming Libraries}

%
NVIDIA's CUTLASS~\cite{cutlass} is a template-based library that
allows users to mix-and-match strategies for developing high-performance
linear algebra algorithms on GPUs.
When a user's program can be expressed with CUTLASS templates, the library
allows for very little code to be modified.
However, when the user's computation departs from a GEMM-like loop 
(such as Flash Attention), the user is responsible for details
like explicit communication, synchronization and warp specialization.
CuTe~\cite{cute} is a subsystem of CUTLASS that provides sophisticated
infrastructure to model complex data layouts and dispatch to specialized assembly
instruction variants.
\name{}'s partitioning operators encapsulate common partitioning patterns
used within CuTe, and leverage CuTe's layout algebra to model data layout
transformations internally.
CuTe is also used in \name{}'s generated code, simplifying the mechanism of
generating architecture specific code.
%
%
The Graphene~\cite{graphene} DSL allows for representing complex data-to-thread 
mappings for matrix assembly instructions, but does not have support
for asynchrony, a core concept in Hopper programming.
ThunderKittens~\cite{thunderkittens} is a new library for Hopper programming
that provides a concise API to target the Tensor Core and TMA.
Like CUTLASS, ThunderKittens places the burden of managing synchronization
and communication on the programmer.
%

\paragraph{Tile-Based Programming Models}

Tile-based programming models like Triton~\cite{triton}
and Pallas~\cite{pallas} have been enormously successful in simplifying the
development of high-performance linear algebra.
These models ask the programmer to express how the computation should be decomposed
onto individual thread blocks of the GPU, and then let the compiler handle the further
decomposition of the block-level program onto the threads.
Our evaluation (\Cref{sec:evaluation}) shows that as the underlying architecture becomes
more complex and as programs diversify, relying on the compiler to make all the decisions
about lowering the block-level program can yield sub-optimal performance.
\name{} recursively applies this key idea of Triton to allow the programmer
to express compute and data decomposition at each level of the machine hierarchy.
\name{}'s separate mapping specification then gives the programmer
control over performance, avoiding heuristics.
At the same time, \name{} provides automation of data movement and 
synchronization, allowing the programming model to remain at a high level.

\paragraph{Functional GPU Programming Models}
Functional approaches to GPU programming have explored using rewrite rules
to capture the hierarchy in efficient GPU programming.
Rasch uses multi-dimensional homomorphisms to express hierarchical decompositions
of data parallel programs targeting the GPU~\cite{derecomp}.
The RISE/ELEVATE~\cite{lift, rise} line of work defines GPU programs as functional
compositions of data parallel operators, and uses a functional rewrite language
to lower the high-level programs into low-level code.
The RISE compiler was also extended to support Turing generation Tensor Cores~\cite{rise-tensor-core}
by exposing the Tensor Core APIs to the functional rewrite primitives.
While these approaches have been shown to achieve high performance,
they have not yet demonstrated the use of asynchronous accelerators 
like the TMA and Hopper Tensor Core.

\paragraph{Scheduling-Based DSL Compilers}
Domain specific languages for image processing~\cite{halide} and
tensor algebra~\cite{taco, tvm} separate the description of the
computation from the optimization strategy, called a \emph{schedule}.
Languages like FireIron~\cite{fireiron} have described how 
to represent optimizations like swizzling and specialized instruction
dispatch in scheduling languages.
The Exo~\cite{exo} language allows for the programmer to 
define custom instructions and memories, and exposes
compositional scheduling primitives to map computation
and data onto the target processors.
%
%
Scheduling languages tend to focus on loop transformations for
nested loop programs. 
\name{} targets a lower-level representation and
automates the management of asynchronous fixed-function units.
DSL compilers could conceivably target \name{} to simplify
the analysis required to target Hopper.

\IGNORE{
TODO (rohany): I tried to find a way to connect this back to this work,
but couldn't come up with a reasonable link here. Our GPUs are really CISC-y
machines, not VLIW. So talking about VLIW is kind of out there without a direct
transition. I don't think there's something super fundamental about compiling
for CISC, it is basically either 1) more instruction selection stuff to find
good instructions or 2) entirely different ways of programming to use the
macro instructions once the CISC stuff gets too crazy.

\paragraph{VLIW Compilation}
\TODO{Ask alex for references}
}

\section{Conclusion}\label{sec:conclusion}

To meet demands for increased performance across different application domains,
processors have and will continue to exploit heterogeneity.
As asynchronous fixed function units are developed and deployed within
processors---NVIDIA's Tensor Cores are a notable example---the familiar bulk-synchronous idioms
used to program these processors are insufficient for some kernels.
In order to maintain control of the complexity in programming processors
with asynchronous fixed function units, we have proposed the \name{} programming
model and compiler.
\name{}'s programming model allows for programmers to describe their logical
computation in a sequential model, and \name{}'s compiler automates the management of asynchronous data movement and matrix-multiplication units.
We demonstrated that kernels generated with \name{} can achieve competitive performance
with hand-tuned implementations while also automating the tedious aspects of achieving high performance.


\section{Acknowledgements}

We thank Vijay Thakkar, Cris Cecka, Pradeep Ramani, 
Balaji Atukuri, Jun Lim for their help with CuTe, CUTLASS and Hopper
programming in general.
We thank Benjamin Driscoll for help with developing the
formalism in this work.
We thank Sean Treichler, Wonchan Lee, Maryam Dehnavi and Fredrik Kjolstad
for helpful discussions about the work.
We thank
Chris Gyurgyik, AJ Root, Marco Siracusa, Olivia Hsu, Bobby Yan, Bala Vinaithirthan, James Dong, and Shiv Sundram
for feedback on this manuscript.


\bibliography{main}

\end{document}

%% file: fe-syntax.tex
\scriptsize
\[
\begin{array}{rlcl} 
    \textsf{Variable} & x & & \\
    \textsf{Constant} & c & & \\
    \textsf{Type} & \tau & \bnfdef & \textsf{int} \bnfalt \textsf{float} \bnfalt
                            \textsf{tensor}(\textsf{int}, \tau) \\ 
    \textsf{Partition Kind} & pk & \bnfdef & \textsf{blocks} \bnfalt \textsf{mma} \\
    
    \textsf{Expression} & e & \bnfdef & x \bnfalt c \bnfalt e \oplus e \bnfalt 
        \textsf{partition}(pk, \overline{e}) \bnfalt e[\overline{e}]   \\ 


    \textsf{Statement} & s & \bnfdef & x = e \bnfalt \textsf{call-external}(f, \overline{e}) \bnfalt \\
    &&& x = \textsf{tunable}(\tau)
    \bnfalt \textsf{launch}(T, \overline{e}) \bnfalt \\
    &&& \textsf{for}~x~\textsf{in}~\textsf{srange}(e, e)~\textsf{launch}(T, \overline{e}) \bnfalt \\
    &&& \textsf{for}~x~\textsf{in}~\textsf{prange}(e, e)~\textsf{launch}(T, \overline{e}) \\
    

    \textsf{Task Name} & T & & \\
    \textsf{Privilege} & pr & \bnfdef & \textsf{read} \bnfalt \textsf{write} \bnfalt \textsf{read-write} \\
    \textsf{Task Variant Kind} & k & \bnfdef & \textsf{inner} \bnfalt \textsf{leaf} \\
    \textsf{Task Variant} & TV & \bnfdef & \textsf{def}~f@\{T,k\}(\overline{(x, \tau, p)})~\textsf{do}~\overline{s}\\ 
    \textsf{Logical Description} & L & \bnfdef & \overline{TV} \\

    \\

    \textsf{Processor} & p & \bnfdef & \textsf{host} \bnfalt \textsf{block} \bnfalt \textsf{warpgroup} \bnfalt \textsf{warp} \bnfalt \textsf{thread} \\
    \textsf{Memory} & m & \bnfdef & \textsf{none} \bnfalt \textsf{global} \bnfalt \textsf{shared} \bnfalt \textsf{register} \\
    \textsf{Tunable Assignment} & ta & \bnfdef & x \mapsto c \\
    \textsf{Task Mapping} & TM & \bnfdef & \textsf{task-mapping}(f, p, \overline{m}, \overline{ta}, \overline{TM}) \\
    \textsf{Mapping Specification} & M & \bnfdef & \overline{TM} \\
\end{array}
\]

%% file: gemm-example.tex
\begin{figure}
\begin{subfigure}[h]{0.62\textwidth}
\begin{lstlisting}
@task("gemm", Inner, reads=["A","B"], writes=["C"])(*\label{fig:cypressgemm:privs}*)
def gemm_host(C: tensor[2,f16],(*\tikzmark{targstart}*)
              A: tensor[2,f16],
              B: tensor[2,f16]):(*\tikzmark{targend}*)
  U, V = tunable(int), tunable(int)(*\tikzmark{ttunablestart}\tikzmark{ttunableend}*)
  # Input tensor sizes are dynamic values.
  M, N, K = C.shape[0], C.shape[1], A.shape[1]
  # Partition C into UxV tiles, and describe 
  # the corresponding rows and columns of A and B.
  Cp = partition_by_blocks(C, (U, V))(*\label{fig:cypressgemm:hostpart}*)
  Ap = partition_by_blocks(A, (U, K))
  Bp = partition_by_blocks(B, (K, V))
  # Launch a parallel group of tasks over each tile.
  for i, j in prange(cdiv(M, U), cdiv(N, V)):(*\tikzmark{tlaunchstart}*)
    launch("gemm", Cp[i, j], Ap[i, 0], Bp[0, j])(*\label{fig:cypressgemm:reclaunch}\tikzmark{tlaunchend}*)

@task("gemm", Inner, reads=["A","B"], writes=["C"])
def gemm_block(C: tensor[2,f16],
               A: tensor[2,f16],
               B: tensor[2,f16]):
  W = tunable(int)
  M, N, K = C.shape[0], C.shape[1], A.shape[1]
  # Break A and B into tiles of width W.
  Ap = partition_by_blocks(A, (M, W))
  Bp = partition_by_blocks(B, (W, N))
  Cacc: tensor[2, f16] = make_tensor(M, N)
  launch("clear", Cacc)(*\label{fig:cypressgemm:inlinelaunch}*)
  # Launch a sequential group of tasks over each tile.
  for k in srange(cdiv(K, W)):
    launch("gemm", Cacc, Ap[0, k], Bp[k, 0])(*\label{fig:cypressgemm:srange}*)
  launch("copy", Cacc, C)

@task("gemm", Inner, reads=["A","B","C"], writes["C"])
def gemm_tile(C: tensor[2,f16],
              A: tensor[2,f16],
              B: tensor[2,f16]):
  WGS = tunable(int)
  M, N, K = C.shape[0], C.shape[1], A.shape[1]
  # Partition the rows for each warpgroup.
  Cp = partition_by_blocks(C, (M/WGS, N))
  Ap = partition_by_blocks(A, (M/WGS, K))
  for i in prange(WGS):
    launch("gemm", Cp[i, 0], Ap[i, 0], B)

@task("gemm", Inner, reads=["A","B","C"], writes["C"])
def gemm_inner(C: tensor[2,f16],
               A: tensor[2,f16],
               B: tensor[2,f16]):
  PIECES, PROC = tunable(int), tunable(processor)
  # MMA partitioning depends on processor and operand position.
  Cp = partition_by_mma(C, WGMMA_64x256x16(), PROC, "C")
  Ap = partition_by_mma(A, WGMMA_64x256x16(), PROC, "A")
  Bp = partition_by_mma(B, WGMMA_64x256x16(), PROC, "B")
  for i in prange(PIECES):
    launch("gemm", Cp[i], Ap[i], Bp[i])

@task("gemm", Leaf, reads=["A","B","C"], writes["C"])
def gemm_thread(C: tensor[2,f16],
                A: tensor[2,f16],
                B: tensor[2,f16]):
  # Dispatch to the PTX WGMMA instruction with CuTe.
  CuTe_warpgroup_gemm(WGMMA_64x256x16(), C, A, B)
  
# Task trees for the `fill` and `copy` tasks elided.
\end{lstlisting}
\caption{Logical program description.}
\label{fig:cypress-gemm-tasks}
\end{subfigure}\hfill%
\begin{subfigure}[h]{0.37\textwidth}
\begin{lstlisting}
[
  TaskMapping(
    instance="gemm_host",
    variant="gemm_host",(*\tikzmark{margstart}*)
    proc=HOST,
    mems=[GLOBAL, GLOBAL, GLOBAL],(*\tikzmark{margend}*)
    tunables={"U": 256, (*\tikzmark{mtunablestart}*)
              "V": 256},(*\tikzmark{mtunablend}*)
    entrypoint=True,
    calls=["gemm_block"](*\tikzmark{mlaunchstart}\tikzmark{mlaunchend}*)
  ),
  TaskMapping(
    instance="gemm_block",(*\tikzmark{mrefinstance}*)
    variant="gemm_block",
    proc=BLOCK,
    mems=[GLOBAL, GLOBAL, GLOBAL](*\label{fig:cypressmapping:globalmap}*),
    tunables={"W": 64},
    calls=["clear_tile",
           "gemm_tile",(*\label{fig:cypressmapping:gemmtiledispatch}*)
           "copy_tile"],
    warpspecialize=True,(*\label{fig:cypressmapping:knob-ws}*)
    pipeline=3(*\label{fig:cypressmapping:knob-pipeline}*)
  ),
  TaskMapping(
    instance="gemm_tile",
    variant="gemm_tile",
    proc=BLOCK,
    mems=[NONE, SHARED, SHARED],(*\label{fig:cypressmapping:nonemap}*)
    tunables={"WGS": 2},
    calls=["gemm_warpgroup"]
  ),
  TaskMapping((*\label{fig:cypressmapping:inst1}*)
    instance="gemm_warpgroup",
    variant="gemm_inner",
    proc=WARPGROUP,
    mems=[NONE, SHARED, SHARED],
    tunables={"PIECES": 4,
              "PROC": WARP}
    calls=["gemm_warp"]
  ),
  TaskMapping((*\label{fig:cypressmapping:inst2}*)
    instance="gemm_warp",
    variant="gemm_inner",
    proc=WARP,
    mems=[NONE, SHARED, SHARED],
    tunables={"PIECES": 32,
              "PROC": THREAD}
    calls=["gemm_thread"]
  ),
  TaskMapping(
    instance="gemm_thread",
    variant="gemm_thread",
    proc=THREAD,
    mems=[REGISTER, SHARED, SHARED],
  ),
  # Mappings for `fill` and
  # `copy` task trees elided.
]
\end{lstlisting}
\caption{Mapping specification.}
\label{fig:cypress-gemm-mapping}
\end{subfigure}
\caption{H100 GEMM implementation developed in \name{}. The logical description expresses the decomposition of computation and data, and the mapping binds tasks and tensors to physical processors and memories.
Numbered bracket annotations indicate components of the logical description controlled by the
mapping specification.
Communication and synchronization are notably absent from the logical description.
}
\label{fig:cypress-gemm}

\begin{tikzpicture}[overlay, remember picture]
    \drawBrace[0.3em]{targstart}{targend}{(1)};
    \drawBrace[0.3em]{margstart}{margend}{(1)};
    \drawBrace[0.3em]{ttunablestart}{ttunableend}{(2)};
    \drawBrace[0.3em]{mtunablestart}{mtunablend}{(2)};
    \drawBrace[0.3em]{tlaunchstart}{tlaunchend}{(3)};
    \drawBrace[0.3em]{mlaunchstart}{mlaunchend}{(3)};

    \draw [->,mygray,thick] (mlaunchend.south) to [bend left] node {} (mrefinstance.east);
    
\end{tikzpicture}

\end{figure}


%% file: dep-analysis.tex
\begin{figure}
\begin{subfigure}[h]{0.5\textwidth}
\begin{lstlisting}
@task("gemm", Inner, reads=["A","B"], writes=["C"])
def gemm_block(C: tensor[2,f16],
               A: tensor[2,f16],
               B: tensor[2,f16]):
  W = tunable(int)
  M, N, K = C.shape[0], C.shape[1], A.shape[1]
  # Break A and B into tiles of width W.
  Ap = partition_by_blocks(A, (M, W))
  Bp = partition_by_blocks(B, (W, N))
  Cacc: tensor[2, f16] = make_tensor(M, N)
  launch("clear", Cacc)
  for k in srange(cdiv(K, W)):
    launch("gemm", Cacc, Ap[0, k], Bp[k, 0])
  launch("copy", Cacc, C)

@task("clear", Inner, writes=["C"])
def clear_inner(C: tensor[2,f16]):
  PIECES = tunable(int)
  PROC = tunable(processor)
  Cp = partition_by_mma(
    C, WGMMA_SM90_64x256x16(), PROC, "C")
  for i in prange(PIECES):
    launch("clear", Cp[i])
\end{lstlisting}
\caption{Candidate GEMM tasks for lowering.}
\end{subfigure}\hfill%
\begin{subfigure}[h]{0.48\textwidth}
\begin{lstlisting}
Ap, Bp = partition(A), partition(B)
C = tensor([M, N], NONE)
C1 = tensor([M, N], NONE)
C1p = partition(C1)
e1: [(4, WARP)] = pfor i in [0, 4), {} do(*\label{fig:dep-analysis-clearpfor1}*)
 CW = tensor([M/4, N], NONE)
 CWp = partition(CW)
 e2: [(32, THREAD)] = pfor j in [0, 32], {} do(*\label{fig:dep-analysis-clearpfor2}*)
  # Concrete shape of register fragment elided.
  CR = tensor([...], RMEM)
  e3: () = call(clear_thread, CR), {}
  e4: () = copy(CR, CWp[j]), {e3}(*\label{fig:dep-analysis-copyout1}*)
  yield e4
 e5: () = copy(CW, C1p[i]), {e2[:]}(*\label{fig:dep-analysis-copyout2}*)
 yield e5
e6: () = copy(C1, C), {e1[:]}
e7: () = for k in [0, cdiv(K, W)), {e6} do
 C2 = tensor([M, N], NONE)
 Ak = tensor([M, W], SMEM)
 Bk = tensor([W, N], SMEM)
 e8: () = copy(C, C1), {}
 e9: () = copy(Ap[0, k], Ak), {}
 e10: () = copy(Bp[k, 0], Bk), {}
 # Recursive lowering of gemm elided.
 e11: () = gemm(C2, Ak, Bk), {e8, e9, e10}
 e12: () = copy(C2, C), {e11}
 yield e12
# Lowering of "copy" task elided.
\end{lstlisting}
\caption{Partial dependence analysis result IR.}
\label{fig:dep-analysis-example-post}
\end{subfigure}
\caption{\name{}'s dependence analysis lowers the task-based logical description into a dependence graph.}
\label{fig:dep-analysis-example}
\end{figure}

%% file: vectorization.tex
\begin{figure}
\begin{subfigure}[h]{0.32\textwidth}
\begin{lstlisting}
C = tensor([M, N], NONE)
C1 = tensor([M, N], NONE)
C1p = partition(C1)
e1: [(4,WARP)] =
pfor i in [0, 4), {} do
 CW = tensor([M/4, N], NONE)
 CWp = partition(CW)
 e2: [(32,THREAD)] =
 pfor j in [0, 32], {} do
  CR = tensor([...], RMEM)
  e3: () =
   call(clear_thread, CR), {}
  e4: () =
   copy(CR, CWp[j]), {e3}
  yield e4
 e5: () =
  copy(CW, C1p[i]), {e2[:]}
 yield e5
e6: () = copy(C1, C), {e1[:]}
\end{lstlisting}
\end{subfigure}\hfill%
\begin{subfigure}[h]{0.32\textwidth}
\begin{lstlisting}
C = tensor([M, N], NONE)
C1 = tensor([M, N], NONE)
C1p = partition(C1)
e1: [(4,WARP)] =
pfor i in [0, 4), {} do
 CW = tensor([M/4, N], NONE)
 CWp = partition(CW)
 j = thread_id()
 CR = tensor([...], RMEM)
 e3: [(32,THREAD)] = 
  call(clear_thread, CR), {}
 e4: [(32,THREAD)] =
  copy(CR, CWp[j]), {e3[j]}
 e5: () =
  copy(CW, C1p[i]), {e4[:]}
 yield e5
e6: () = copy(C1, C), {e1[:]}
\end{lstlisting}
\end{subfigure}\hfill%
\begin{subfigure}[h]{0.32\textwidth}
 \begin{lstlisting}
C = tensor([M, N], NONE)
C1 = tensor([M, N], NONE)
C1p = partition(C1)
i = warp_id()
CW = tensor([M/4, N], NONE)
CWp = partition(CW)
j = thread_id()
CR = tensor([...], RMEM)
e3: [(4,WARP),(32,THREAD)] =
 call(clear_thread, CR), {}
e4: [(4,WARP),(32,THREAD)] =
 copy(CR, CWp[j]), {e3[i, j]}(*\label{fig:vec-pointwise-event}*)
e5: [(4,WARP)] =
 copy(CW, C1p[i]), {e4[i, :]}(*\label{fig:vec-bcast-event}*)
e6: () = copy(C1, C), {e5[:]}
\end{lstlisting}
\end{subfigure}

\begin{subfigure}[b]{0.32\textwidth}
\caption{Original IR (\Cref{fig:dep-analysis-example}).} 
\end{subfigure}\hfill%
\begin{subfigure}[b]{0.32\textwidth}
\caption{Vectorized inner loop.}
\end{subfigure}\hfill%
\begin{subfigure}[b]{0.32\textwidth}
\caption{Both loops vectorized.}
\label{fig:vec-final}
\end{subfigure}
\caption{Example of \name{}'s vectorization pass flattening implicit loops. Event arrays capture dependencies.}
\label{fig:vectorization}
\end{figure}

%% file: copy-elim.tex
\begin{figure}
\begin{subfigure}[h]{0.24\textwidth}
\begin{lstlisting}
b1
e2: [(N, PROC)] =
 copy(t, P[i]), {e1}
e3: [(N, PROC)] =
 copy(P[i], t), {e2[:]}
b2
\end{lstlisting}
\end{subfigure}\hfill%
\begin{subfigure}[h]{0.24\textwidth}
\begin{lstlisting}
e2: () = for i, {e1} do
 e3: et1 = copy(P[j], t)
 b
 e5: et1 = 
   copy(t, P[j]), {e4}
 yield e5
\end{lstlisting}
\end{subfigure}\hfill%
\begin{subfigure}[h]{0.24\textwidth}
\begin{lstlisting}
e1: et = copy(P[i], t)
b1
e2: et = copy(P[i], t)
b2
\end{lstlisting}
\end{subfigure}\hfill%
\begin{subfigure}[h]{0.24\textwidth}
\begin{lstlisting}
b1
e2: et = 
 copy(t, t), {e1}
b2
\end{lstlisting}
\end{subfigure}

\begin{subfigure}[t]{0.24\textwidth}
$\quad\quad\quad\quad\boldsymbol{\downarrow}$
\end{subfigure}\hfill%
\begin{subfigure}[t]{0.24\textwidth}
$\quad\quad\quad\quad\boldsymbol{\downarrow}$
\end{subfigure}\hfill%
\begin{subfigure}[t]{0.24\textwidth}
$\quad\quad\quad\quad\boldsymbol{\downarrow}$
\end{subfigure}\hfill%
\begin{subfigure}[t]{0.24\textwidth}
$\quad\quad\quad\quad\boldsymbol{\downarrow}$
\end{subfigure}

\begin{subfigure}[h]{0.24\textwidth}
\begin{lstlisting}
b1
b2 [e3 / e1]
\end{lstlisting}
\end{subfigure}\hfill%
\begin{subfigure}[h]{0.24\textwidth}
\begin{lstlisting}
e3 = copy(P[j], t), {e1}
e2 = for i, {e3} do
 b
 yield e4
e5 = copy(t, P[j]), {e2}
\end{lstlisting}
\end{subfigure}\hfill%
\begin{subfigure}[h]{0.24\textwidth}
\begin{lstlisting}
e1 = copy(P[i], t)
b1
b2[e2 / e1]
\end{lstlisting}
\end{subfigure}\hfill%
\begin{subfigure}[h]{0.24\textwidth}
\begin{lstlisting}
b1
b2[e2 / e1]
\end{lstlisting}
\end{subfigure}

\begin{subfigure}[b]{0.24\textwidth}
\caption{Spill elimination}
\label{fig:spill-elimination}
\end{subfigure}\hfill%
\begin{subfigure}[b]{0.24\textwidth}
\caption{Spill Hoisting}
\label{fig:spill-hoisting}
\end{subfigure}\hfill%
\begin{subfigure}[b]{0.24\textwidth}
\caption{Duplicate Elimination}
\label{fig:dup-copy-elim}
\end{subfigure}\hfill%
\begin{subfigure}[b]{0.24\textwidth}
\caption{Self Copy Elimination}
\label{fig:self-copy-elim}
\end{subfigure}

\caption{Selected set of copy elimination patterns. We let \texttt{b*} refer to blocks of operations. The \texttt{b[e2 / e1]} syntax indicates that we substitute \texttt{e2} with \texttt{e1} inside of the block.}
\label{fig:copy-elim-patterns}
\end{figure}